\documentclass[final,authoryear,11pt]{elsarticle} 
\journal{Ocean Engineering}
	
\usepackage{amsmath,amssymb,amsfonts}
\usepackage{tikz}
\usetikzlibrary{arrows}
\usepackage{graphicx}

\newcommand{\pd}[1]{f\left({} #1 \right){}}
\newcommand{\T}{^{\text{T}}}
\newcommand{\gpdt}[1]{y_{#1}}
\newcommand{\gpnd}{N} 
\newcommand{\gpcv}[1]{\theta_{#1}}
\newcommand{\gpdm}{\Theta{}} 
\newcommand{\gpsh}[1]{\xi_{#1}}
\newcommand{\gpshf}[1]{\xi{}\left({} #1 \right){}} 
\newcommand{\gpsc}[1]{\sigma_{#1}}
\newcommand{\gpscf}[1]{\sigma{}\left({} #1 \right){}} 
\newcommand{\gpnu}[1]{\nu_{#1}}
\newcommand{\gpnuf}[1]{\nu{}\left({} #1 \right){}}
\newcommand{\gpth}[1]{\mu_{#1}}
\newcommand{\gpthf}[1]{\mu{}\left({} #1 \right){}} 
\newcommand{\gprt}[1]{\rho_{#1}}
\newcommand{\gprtf}[1]{\rho{}\left({} #1 \right){}} 
\newcommand{\gpr}{\eta{}}
\newcommand{\gprf}[1]{\eta{}\left({} #1 \right){}} 
\newcommand{\prbs}[2]{B_{#1}\left({} #2 \right){}}
\newcommand{\prbsn}[1]{B_{#1}} 
\newcommand{\prnbs}[1]{p_{#1}} 
\newcommand{\prp}[1]{\beta_{#1}}
\newcommand{\prpn}[1]{\lambda_{#1}}
\newcommand{\prcr}[1]{Q_{#1}}
\newcommand{\rgh}[1]{R_{#1}}
\newcommand{\prpna}[1]{a_{#1}}
\newcommand{\prpnb}[1]{b_{#1}}
\newcommand{\spbf}[2]{b_{#1}\left({} #2 \right){}}
\newcommand{\frnp}{n_{p}}
\newcommand{\frccos}[1]{a_{#1}}
\newcommand{\frcsin}[1]{b_{#1}}

\newcommand{\gcnp}{\prnbs{}}
\newcommand{\gckn}[1]{\hat{\theta}_{#1}}
\newcommand{\gcif}[2]{I_{#1}\left({} #2 \right){}}
\newcommand{\gccv}[1]{V_{#1}}
\newcommand{\gccf}[1]{r_{#1}}
\newcommand{\Dks}[1]{D_{\text{ks}}\left({} #1 \right){}}
\newcommand{\Dcvm}[1]{D_{\text{cm}}\left({} #1 \right){}}
\newcommand{\Dkl}[1]{D_{\text{kl}}\left({} #1 \right){}}
\newcommand{\DF}[2]{F_{#1}\left({} #2 \right){}}
\newcommand{\DFna}[1]{F_{#1}}
\newcommand{\Df}[2]{f_{#1}\left({} #2 \right){}}
\newcommand{\mhar}[1]{A\left({} #1 \right){}}
\newcommand{\mlpr}[1]{\Omega_{#1}}
\newcommand{\mlhpr}{\Gamma{}}
\newcommand{\pexc}{\neg{}}
\newcommand{\mlsm}[2]{\beta_{#1}^{(#2)}}
\newcommand{\mlnw}[1]{\beta_{#1}^{*}}
\newcommand{\mlep}[1]{\nu_{#1}}
\newcommand{\mllk}[1]{L\left({} #1 \right){}}
\newcommand{\mllks}[1]{L^*\left({} #1 \right){}}
\newcommand{\mlgr}[1]{D\left({} #1 \right){}}
\newcommand{\mlhs}[2]{G^{#1}\left({} #2 \right){}}
\newcommand{\gba}[1]{\hat{a}_{#1}}
\newcommand{\gbb}[1]{\hat{b}_{#1}}
\newcommand{\irgr}[1]{V\left({} #1 \right){}}
\newcommand{\irhs}[1]{W\left({} #1 \right){}}

\begin{document}

	
\begin{frontmatter}
\title{Statistics of extreme ocean environments: Non-stationary inference for directionality and other covariate effects}
\author[dur]{Matthew Jones}
\author[man]{David Randell}
\author[sar]{Kevin Ewans}
\author[man]{Philip Jonathan\corref{cor1}}
\cortext[cor1]{Corresponding author. Email: {\tt philip.jonathan@shell.com}}
\address[dur]{Department of Mathematical Sciences, Durham University, Durham DH1 3LE, United Kingdom.}
\address[man]{Shell Projects \& Technology, Manchester M22 0RR, United Kingdom.}
\address[sar]{Sarawak Shell Bhd., 50450 Kuala Lumpur, Malaysia.}


\begin{abstract}
Numerous approaches are proposed in the literature for non-stationarity marginal extreme value inference, including different model parameterisations with respect to covariate, and different inference schemes. The objective of this article is to compare some of these procedures critically. We generate sample realisations from generalised Pareto distributions, the parameters of which are smooth functions of a single smooth periodic covariate, specified to reflect the characteristics of actual samples from the tail of the distribution of significant wave height with direction, considered in the literature in the recent past. We estimate extreme values models (a) using Constant, Fourier, B-spline and Gaussian Process parameterisations for the functional forms of generalised Pareto shape and (adjusted) scale with respect to covariate and (b) maximum likelihood and Bayesian inference procedures. We evaluate the relative quality of inferences by estimating return value distributions for the response corresponding to a time period of $10 \times$ the (assumed) period of the original sample, and compare estimated return values distributions with the truth using Kullback-Leibler, Cramer-von Mises and Kolmogorov-Smirnov statistics. We find that Spline and Gaussian Process parameterisations estimated by Markov chain Monte Carlo inference using the mMALA algorithm, perform equally well in terms of quality of inference and computational efficiency, and generally perform better than alternatives in those respects.
\end{abstract}
\end{frontmatter}


\section{Introduction}\label{Sct1:Int}
%
Accurate estimates of the likely extreme environmental loading on an offshore facility are vital to enable a design that ensures the facility is both structurally reliable and economic. This involves estimating the extreme value behaviour of meteorological and oceanographic (metocean) variables that quantify the various environmental loading quantities, primarily winds, wave, and currents. Examples of such parameters are significant wave height, mean wind speed and mean current speed. These characterise the environment for a given short period of time within which the environment is assumed to be stationary.

The long-term variability of these parameters is however non-stationary, in particular with respect to time, space and direction. From a temporal point of view metocean parameters generally have a strong seasonal variation, with an annual periodicity, and longer term variations due to decadal or semi-decadal climate variations. At any given location, the variability of a particular parameter is also dependent on the direction; for example, wind forcing is typically stronger from some directions than others, and fetch and water depth effects can strongly influence the resulting magnitude of the waves. Clearly these effects will vary with location: a more exposed location will be associated with longer fetches, resulting in a more extreme wave climate. \par

When estimating the long-term variability of parameters, such as significant wave height, the non-stationary effects associated with e.g. direction and season can be incorporated by treating direction and season as covariates. The common practice is to perform extreme value analysis of hindcast data sets, which include many years of metocean parameters, along with their associated covariates. Such data sets have all the information needed for input to covariate analysis.\par

From a design perspective, the metocean engineer is often required to specify return values for directional sectors such as octants centred on the cardinal and semi-cardinal directions. These directional return value estimates must be consistent with the estimated omnidirectional return value. In a similar manner, return values may be required corresponding to particular seasons or months of the year, consistent with an all-year return value. Clearly, therefore, efficient and reliable inference for non-stationary extremes is of considerable practical interest, requiring estimation of (a) the rate and (b) the size of rare events. This work addresses the latter of these objectives.

A non-stationary extreme value model is generally superior to the alternative ``partitioning'' method sometimes used within the ocean engineering community. In the partitioning method, the sample is partitioned into subsets corresponding to approximately constant values of covariate(s); independent extreme value analysis is then performed on each subset. For example, in the current work we might choose to partition the sample into directional octants, and then estimate (8 independent stationary) extreme value models for each of the octants. There are two main reasons for favouring a non-stationarity model over the partitioning method. Firstly, the partitioning approach incurs a loss in statistical efficiency of estimation, since parameter estimates for subsets with similar covariate values are estimated independently of one another, even though physical insight would require parameter estimates to be similar.  This problem worsens as the number of covariates and covariate subsets increases, and the sample size per subset decreases as a result. In the non-stationary model, we require that parameter estimates corresponding to similar values of covariates be similar, and optimise the degree of similarity during inference. For this reason, parameter uncertainty from the non-stationary model is generally smaller than from the partitioning approach. Secondly, the partitioning approach assumes that, within each subset, the sub-sample for extreme value modelling is homogeneous with respect to covariates. In general it is difficult to estimate what effect this assumption might have on parameter and return value estimates (especially when large intervals of values of covariates are combined into a subset). In the non-stationary model, we avoid the need to make this assumption.

Numerous articles have reported the essential features of extreme value analysis (e.g. \citealt{DvsSmt90}) and the importance of considering different aspects of covariate effects (e.g. \citealt{NrtJntRnd15}). \cite{CrtChl81} considers estimation of annual maxima from monthly data, when the distribution functions of monthly extremes are known.  \cite{ClsWls94} describes directional modelling of extreme wind speeds using a Fourier parameterisation. \cite{SctGds00} models the long-term time series of significant wave height with non-linear threshold models. \cite{AndCrt01} reports that estimates for 100-year significant wave height from an extreme value model ignoring seasonality are considerably smaller than those obtained using a number of different seasonal extreme value models. \cite{ChvDmlEmb04} describes smooth extreme value models in finance and insurance. \cite{ChvDvs05} provides a straight-forward description of a nonhomogeneous Poisson model in which occurrence rates and extreme value properties are modelled as functions of covariates. \cite{ColEA06} uses a Bayesian hierarchical model to characterise extremes of lichen growth. \cite{RnrEA06} considers identification of changes in peaks over threshold using Bayesian inference. \cite{FctWls06} uses a hierarchical model to identify location and seasonal effects in marginal densities of hourly maxima for wind speed. \cite{MndEA08} considers seasonal non-stationarity in extremes of NOAA buoy records. \cite{RndEA15} discusses estimation for return values for significant wave height in the South China Sea using a directional-seasonal extreme value model.\par

\begin{figure}
\centering
\includegraphics[width=1.0\columnwidth]{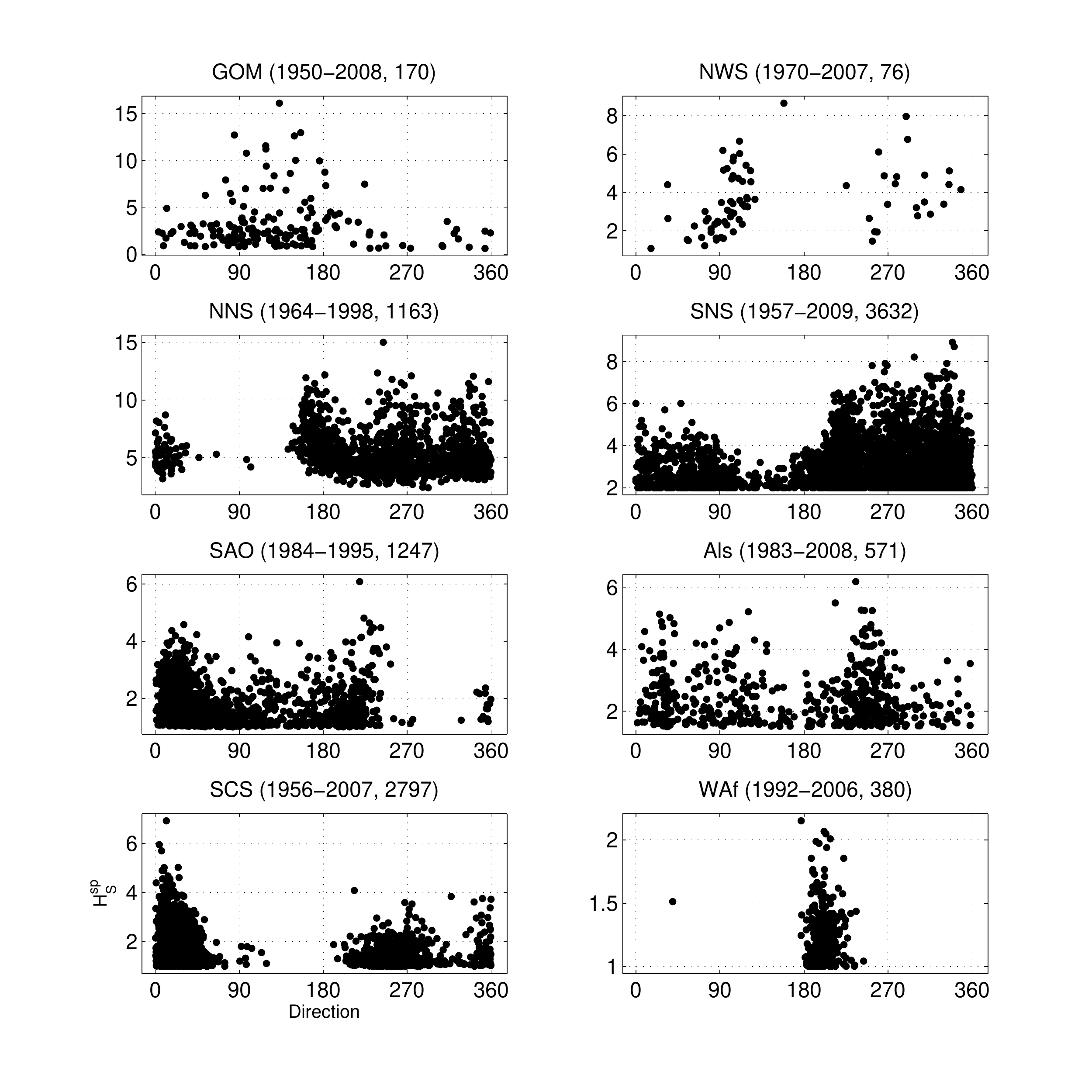}
\caption{Hindcast storm peak significant wave height on direction for 8 locations worldwide. From right to left, top to bottom: Gulf of Mexico (GOM), North-West Shelf of Australia (NWS), Northern North Sea (NNS), Southern North Sea (SNS), South Atlantic Ocean (SAO), Alaska (Als), South China Sea (SCS) and West Africa (WAf). Panel titles give the location, the sample period and storm peak sample size. Please refer to \cite{RndZnnVglEA14} for details of data sources. \label{Fgr1}}
\end{figure}

\cite{RndZnnVglEA14} explores the directional characteristics of hindcast storm peak significant wave height with direction for locations in the Gulf of Mexico, North-West Shelf of Australia, Northern North Sea, Southern North Sea, South Atlantic Ocean, Alaska, South China Sea and West Africa. Figure \ref{Fgr1} illustrates the essential features of samples such as these. The rate and magnitude of occurrences of storm events vary considerably between locations, and with direction at each location. There are directional sectors with effectively no occurrences, there is evidence of rapid changes in characteristics with direction and of local stationarity with direction. Any realistic model for such samples needs to be non-stationary with respect to direction.

The objective of this article is to evaluate critically different procedures for estimating non-stationary extreme value models. We quantify the extent to which extreme value analysis of samples of peaks over threshold exhibiting clear non-stationarity with respect to covariates, such as those in Figure \ref{Fgr1} or simulation case studies in Section \ref{Sct3:EvlMth} below, is influenced by a particular choice of model parameterisation or inference method. The 6 simulation case studies introduced in Section \ref{Sct3:EvlMth} are constructed to reflect the general features of the samples in Figure \ref{Fgr1}, with the advantage that the statistical characteristics of the case studies are known exactly, allowing objective evaluation and comparison of competing methods of model parameterisation and inference. Our aim is that the results of this study are generally informative about any application of non-stationary extreme value analysis. We generate sample realisations from generalised Pareto distributions, the parameters of which are smooth functions of a single smooth periodic covariate. Then we estimate extreme value models (a) using Constant, Fourier, B-spline and Gaussian Process parameterisations for the functional forms of generalised Pareto parameters with respect to covariate and (b) maximum likelihood and Bayesian inference procedures. We evaluate the relative quality of inferences by estimating return value distributions for the response corresponding to a time period of $10 \times$ the (assumed) period of the original sample, and compare estimated return values distributions with the truth using Kullback-Leibler (e.g. \citealt{PERE2008}), Cramer-von Mises (e.g. \citealt{And1962}) and Kolmogorov-Smirnov statistics. We cannot hope to compare all possible parameterisations, but choose four parameterisations useful in our experience. Similarly, there are many competing approaches for maximum likelihood and Bayesian inference, and general interest in understanding their relative characteristics. For example, \cite{SmtNlr87} compares maximum likelihood and Bayesian inference for the three-parameter Weibull distribution. In this work, we choose to compare frequentist penalised likelihood maximisation (see Section 2.3) with two Markov chain Monte Carlo (MCMC) methods of different complexities. Non-stationary model estimation is a growing field. There is a huge literature on still further possibilities for parametric (e.g. Chebyshev, Legendre and other polynomial forms) and non-parametric (e.g. Gauss-Markov random fields and radial basis functions) model parameterisations with respect to covariates. Moreover, in extreme value analysis, pre-processing of a response to near stationarity (e.g. using a Box-Cox transformation) is preferred (e.g. 

The outline of the paper is as follows. Section \ref{Sct2:EstNSE} outlines the different model parameterisations and inference schemes under consideration. Section \ref{Sct3:EvlMth} describes underlying model forms used to generate samples for inference, outlines the procedure for estimation of return value distributions and their comparison, and presents results of those comparisons. Section \ref{Sct4:Dsc} provides discussion and conclusions.

\section{Estimating non-stationary extremes} \label{Sct2:EstNSE}

Consider a random variable $Y$ representing an environmental variable of interest such as significant wave height. The characteristics of $Y$ are dependent on covariates such as (wave) direction, season, location and fetch. In this work we assume that a single periodic covariate $\theta$ (typically direction, or season) is sufficient to characterise the non-stationarity of $Y$. That is, we assume that $Y|\theta$ has a stationary distribution. For exceedances $Y-\mu(\theta)$ of some high threshold $\mu(\theta)$, extreme value theory suggest that the conditional distribution of $Y-\mu(\theta)$ given that $Y>\mu(\theta)$ can be approximated by the generalised Pareto distribution

\[\mathrm{Pr} \left( Y>y | Y>\mu(\theta) \right) = \frac{1}{\sigma(\theta)} \left( 1 + \frac{\xi(\theta)}{\sigma(\theta)} \left( y - \mu(\theta) \right) \right)^{-1/\xi(\theta)} \text{   .   }\]

For design purposes, return value distributions are typically estimated using peaks-over-threshold of significant wave height. The characteristics of these peaks, each corresponding to a different storm event, vary typically with respect to wave direction. Conditional on wave direction, the peaks are reasonably assumed to be independent of one another. To evaluate the relative performance of different approaches to non-stationary extreme value analysis, it is therefore natural to use random simulation from generalised Pareto models with known directional characteristics.

\subsection{Generalised Pareto model}\label{Sct2.1:GnrPrt}

We assume we observe a sample $\gpdt{} = \{{}\gpdt{1},\ldots{},\gpdt{\gpnd{}}\}{}$ of peaks over threshold drawn independently from a generalised Pareto distribution, the parameters of which are functions of corresponding observed covariate values $\gpdm{} = \{{}\gpcv{1},\ldots{},\gpcv{\gpnd{}}\}{}$. The sample likelihood is a product of generalised Pareto (GP) likelihoods for each of the observations
\begin{align*}
\pd{\gpdt{}|\gpdm{},\gpsh{},\gpsc{},\gpth{}} & = \prod_{i=1}^{\gpnd{}} \pd{\gpdt{i}|\gpshf{\gpcv{i}},\gpscf{\gpcv{i}},\gpthf{\gpcv{i}}} \nonumber\\
& = \prod_{i=1}^{\gpnd{}} \frac{1}{\gpscf{\gpcv{i}}}\left({}1+\gpshf{\gpcv{i}}\frac{(\gpdt{i}-\gpthf{\gpcv{i}})}{\gpscf{\gpcv{i}}}\right)^{-1/\gpshf{\gpcv{i}}-1} \label{GPLik}
\end{align*}
where $\gpshf{\gpcv{}}$ and $\gpscf{\gpcv{}}$ are the shape and scale parameters as functions of covariate. We do not attempt to estimate the threshold function $\gpthf{\gpcv{}}$, assuming it is $0$ for all covariate values. We also assume that the rate of occurrence $\gprtf{\gpcv{}}$ of exceedances of $\gpth{}$ varies with covariate, but that $\gprtf{\gpcv{}}$ is known (see Section \ref{Sct3.1:Css}). It is computationally advantageous (e.g. \citealt{CoxRed87}, \citealt{ChvDvs05}) to transform variables from $(\gpsh{},\gpsc{})$ to the asymptotically independent pair $(\gpsh{},\gpnu{})$, where $\gpnuf{\gpcv{}} = \gpscf{\gpcv{}}(1+\gpshf{\gpcv{}})$. Inference therefore amounts to estimating the smooth functions $\gpshf{\gpcv{}}$ and $\gpnuf{\gpcv{}}$, although we usually choose to illustrate the analysis in terms of $\gpshf{\gpcv{}}$ and $\gpscf{\gpcv{}}$. In practical application, estimation of $\mu(\theta)$ is itself generally also problematic (e.g. \citealt{ScrMcd12}), particularly in the presence of non-stationarity (\citealt{NrtJnt11}), but as necessary for inference as reliable estimation of $\xi(\theta)$ and $\sigma(\theta)$. We choose to focus on the latter in this work.

\subsection{Covariate parameterisations}\label{Sct2.2:CvrPrm}
%
To accommodate non-stationarity, we parameterise $\gpsh{}$ and $\gpnu{}$ as linear combinations of unknown parameters $\prp{\gpsh{}}$ and $\prp{\gpnu{}}$ respectively, where
\begin{align}
\gpnuf{\gpcv{}} = \prbs{\gpnu{}}{\gpcv{}}\prp{\gpnu{}} \text{ ,  and  } \gpshf{\gpcv{}} = \prbs{\gpsh{}}{\gpcv{}}\prp{\gpsh{}} \nonumber
\end{align}
and $\prbs{\gpnu{}}{\gpcv{}}$ and $\prbs{\gpsh{}}{\gpcv{}}$ are row vectors of basis functions evaluated at $\gpcv{}$. We consider four different forms of basis function, corresponding to Constant (stationary), Fourier, Spline and Gaussian Process parameterisations for $\gpshf{\gpcv{}}$ and $\gpnuf{\gpcv{}}$, as described below.

Physical considerations suggest that we should expect GP model parameters to vary smoothly with covariate. In general, the basis parameterisation introduced here permits estimation of functional forms for GP model parameters which are too variable (or too rough) with respect to covariate. We therefore need a mechanism to restrict the roughness of functional forms, such that their roughness is optimal given the evidence in the data. For each parameterisation, we therefore specify roughness matrices $\prcr{\gpr{}}$ (for $\gpr{} = \gpsh{},\gpnu{}$) to regulate the roughness of $\gprf{\gpcv{}}$ with respect to $\gpcv{}$ during inference. This ensures that the elements of $\prp{\gpr{}}$ weight the individual basis functions in such a way that the resulting estimate is optimally smooth in some sense. The form of the roughness penalty term $\rgh{\gpr{}}$ is $\frac{1}{2}\prpn{\gpr{}}\prp{\gpr{}}' \prcr{\gpr{}} \prp{\gpr{}}$, for some roughness coefficient $\prpn{\gpr{}}$. The penalty is incorporated directly within a penalised likelihood for maximum likelihood inference, and within a prior distribution for $\prp{\gpr{}}$ in Bayesian inference, as described in Section \ref{Sct2.3:Inf}.

\subsubsection*{Constant (stationary) parameterisation}
%
In the Constant parameterisation, the values of $\gpsh{}$ and $\gpnu{}$ do not vary with respect to $\gpcv{}$. We therefore adopt a scalar basis function which is constant across all values of covariate, so that $\prbs{\gpnu{}}{\gpcv{}} = \prbs{\gpsh{}}{\gpcv{}} = 1$, and corresponding roughness matrices $\prcr{\gpnu{}} = \prcr{\gpsh{}} = 1$. We do not expect the return value distributions estimated under this parameterisation to fare well in general in our comparison, since samples are generated from non-stationary distributions. Quality of fit is expected to be poor, at least in some intervals of covariate. However, many practitioners continue to use stationary extreme value models, perhaps with high thresholds to mitigate non-stationarity, in applications; inclusion of a stationary parameterisation provides a useful point of reference for comparison, therefore.

\subsubsection*{Spline parameterisation}
%
Under a Spline parameterisation, the vector of basis functions for each of $\gpnu{}$ and $\gpsh{}$ is made up of $\prnbs{}$ local polynomial B-spline functions with compact support, joined at a series of knots evenly spaced in the covariate domain (e.g. \citealt{ElrMrx10})
\begin{equation*}
\prbs{\gpnu{}}{\gpcv{}} = \prbs{\gpsh{}}{\gpcv{}} = \begin{pmatrix}
\spbf{1}{\gpcv{}} & \cdots{} & \spbf{\prnbs{}}{\gpcv{}}
\end{pmatrix}  \text{   .   }
\end{equation*}
We specify roughness matrices $\prcr{\gpnu{}} = \prcr{\gpsh{}} = D\T{}D$ which penalise squared differences between adjacent elements of the coefficient vectors, where
\begin{equation*}
D = \begin{pmatrix}
-1 & 1 & 0 & \cdots{} & 0 \\
0 & -1 & 1 &  & 0 \\
\vdots{} & & & \ddots{} & \vdots{} \\
0 & 0 & 0 & \cdots{} & 1
\end{pmatrix}
\end{equation*}
is a $(\prnbs{}-1)\times{}\prnbs{}$ difference matrix. In this work we set $p$ to 50.

\subsubsection*{Fourier parameterisation}

We use basis vectors composed of sine and cosine functions of $\frnp{}$ different periods
\begin{equation*}
\prbs{\gpnu{}}{\gpcv{}} = \prbs{\gpsh{}}{\gpcv{}} = \begin{pmatrix}
1 & \sin{}(\gpcv{}) & \sin{}(2\gpcv{}) & \cdots{} & \sin{}(\frnp{}\gpcv{}) & \cos{}(\gpcv{}) & \cdots{} & \cos{}(\frnp{}\gpcv{})
\end{pmatrix}  \text{   .   }
\end{equation*}
The roughness matrix is computed by imposing a condition on the squared second derivative of the resulting parameter function. If we write
\begin{equation*}
\gpr{}(\gpcv{}) = \sum_{k=1}^{\frnp{}} \left( \frccos{\gpr{}k}\cos{}(k\gpcv{}) + \frcsin{\gpr{}k}\sin{}(k\gpcv{}) \right)
\end{equation*}
where $\gpr{} = \gpsh{}$ or $\gpnu{}$, and $\frccos{\gpr{}k}$ and $\frcsin{\gpr{}k}$ are the parameters from $\prp{\gpr{}}$ corresponding respectively to the sine and cosine functions of period $k$. The roughness criterion (from \citealt{JONA2013}) becomes
\begin{equation*}
\rgh{\gpr{}} = \int_{0}^{2\pi{}} (\gpr{}''(\gpcv{}))^{2} d\gpcv{} = \sum_{k=1}^{\frnp{}} k^{4}(\frccos{\gpr{}k}^{2}+\frcsin{\gpr{}k}^{2}) \nonumber
\end{equation*}
such that the penalty matrix can be written in matrix form as
\begin{equation*}
\prcr{\gpr{}} = \text{diag}\left({}0,1,2^{4},\ldots{},k^{4},\ldots{},\frnp{}^{4},1,2^{4},\ldots{},k^{4},\ldots{},\frnp{}^{4}\right){} \nonumber
\end{equation*}
for the $\prnbs{}=2\frnp{}+1$ Fourier parameters $(\frccos{\gpr{}0}, \frccos{\gpr{}1}, \ldots{},\frccos{\gpr{}\frnp{}}, \frcsin{\gpr{}1}, \ldots{},\frcsin{\gpr{}\frnp{}})$. In this work, we set the value of $n_p$ to 25 so that $p=51$.

\subsubsection*{Gaussian Process parameterisation}
%
For a set of $\gcnp{}$ nodes $\{\gckn{1}, \gckn{2}, \ldots, \gckn{p}\}$ on the covariate domain, we use a Gaussian Process parameterisation \citep{RASW2006} , and relate each covariate input to a knot using the following basis vectors
\begin{equation}
\prbs{\gpnu{}}{\gpcv{}} = \prbs{\gpsh{}}{\gpcv{}} = \begin{pmatrix}
\gcif{1}{\gpcv{}} & \cdots{} & \gcif{\gcnp{}}{\gpcv{}}
\end{pmatrix} \nonumber
\end{equation}
where the indicator functions $\gcif{j}{.}$ are defined as
\begin{equation}
\gcif{j}{\gpcv{}} = \begin{cases}
1 & \text{if } |\gpcv{}-\gckn{j}|<|\gpcv{}-\gckn{k}| \text{   } \forall{k\neq{}j} \\
0 & \text{otherwise} \text{   .   }
\end{cases} \nonumber
\end{equation}
Roughness matrices $\prcr{\gpr{}}$ (where $\gpr{} = \gpnu{},\gpsh{}$) are defined by the coefficient correlation matrix $\gccv{\gpr{}}$ via $\prcr{\gpr{}} = \gccv{\gpr{}}^{-1}$, and the elements of $\gccv{\gpr{}}$ generated by a periodic squared exponential covariance function \citep{MACK1998}
\begin{equation}
\gccv{\gpr{}jk} = \exp{}\left({}-\frac{2}{\gccf{\gpr{}}^{2}}\sin{}\left({}\frac{\gckn{j}-\gckn{k}}{2}\right)^{2}\right){} \nonumber
\end{equation}
where $\gccf{\gpr{}}$ are correlation lengths for each of the parameters, fixed to likely values by comparison with the covariate functions used to generate the data; a value of $r_\eta$ for 0.6 was used throughout. $\gccv{\gpr{}}^{-1}$ penalises on the angular difference between the $j^{th}$ and $k^{th}$ nodes, reducing in value from $\exp(0)$ at angular difference zero to $\exp(-2/\gccf{\gpr{}}^2)$ at angular difference 180$^\circ$. Estimating the Gaussian Process parameterisation on the partitioned covariate domain, as opposed to fitting it to each of the data inputs, greatly reduces the number of parameters to estimate, and is physically reasonable provided that $p$ is sufficiently large. In this work, we use $p=50$ equally-spaced nodes. Estimating a parameter for each data point would have made the computational burden for the Gaussian Process parameterisation significantly greater than that for any of the other parameterisations. For example, computations with roughness matrix $\prcr{\gpr{}}$ (of dimension $p \times p$ on the partitioned covariate domain) are more efficient that those using the $N \times N$ version of $\prcr{\gpr{}}$ ($N \approx 1000)$ defined per data input.

\subsubsection*{Model complexity}
%
We assume it is known from physical considerations that the extremal characteristics of the environmental variable of interest (e.g. significant wave height) vary smoothly with directional covariate. Specifically, we expect the form for the tail of the distribution to be homogeneous within a narrow directional sector of width $\approx 10^\circ$. This in turn suggests a suitable minimum complexity for the non-stationary model parameterisations considered in this work. For the Spline parameterisation, we set $p=50$ corresponding to 50 spline basis functions equally spaced on $[0,360)$, with a distance between peaks of adjacent spline basis functions of $7.2^\circ$. We estimate the Gaussian Process on a regular partition of the covariate domain into $p=50$ bins; the distance between the centres of adjacent bins is again $7.2^\circ$. For the Fourier parameterisation, we use a Fourier order of $n_p=25$ (and $p=51$), such that half a wavelength of the highest frequency Fourier component again corresponds to $7.2^\circ$. In this way, we expect the directional resolution of these three model parameterisations to be comparable. Similarly, the number $p$ of basis coefficients to be estimated in each of the three parameterisations is comparable. In this way, we hope to focus fairly in the analysis below on the different inferential challenges presented by Spline, Fourier and Gaussian Process parameterisations and different estimation schemes for problems of comparable complexity. The stationary Constant parameterisation is clearly less complex (with $p=1$), but a useful baseline for comparison: we expect inferences from the Constant parameterisation to be relatively poor, and therefore make obvious the need to consider non-stationarity.

\subsection{Inference procedures}\label{Sct2.3:Inf}
%
We consider two methods for estimating parameters and return value distributions for the models and parameterisations described above, namely (a) maximum penalised likelihood estimation with bootstrapping to quantify uncertainties, and (b) (two forms of) Bayesian inference using Markov Chain Monte Carlo (MCMC). These are discussed below.

\subsubsection*{Maximum likelihood estimation}
%
We use an iterative back-fitting optimisation (see Appendix) to minimise the penalised negative log likelihood $-\mllks{\gpdt{}|\prp{\gpsh{}}, \prp{\gpnu{}};\lambda_{\gpsh{}},\lambda_{\gpnu{}}}$ with respect to $\prp{\gpsh{}}$ and $\prp{\gpnu{}}$ for given roughness coefficients $\lambda_{\gpsh{}}$ and $\lambda_{\gpnu{}}$, where
\begin{eqnarray*}
-\mllks{\gpdt{}|\prp{\gpsh{}}, \prp{\gpnu{}};\lambda_{\gpsh{}},\lambda_{\gpnu{}}} &=& -\mllk{\gpdt{}|\prp{\gpsh{}}, \prp{\gpnu{}}} + \rgh{\gpsh{}} + \rgh{\gpnu{}}\\
&=& -\mllk{\gpdt{}|\prp{\gpsh{}}, \prp{\gpnu{}}} + \frac{1}{2} \prpn{\gpsh{}} \prp{\gpsh{}}' \prcr{\gpsh{}} \prp{\gpsh{}} + \frac{1}{2} \prpn{\gpnu{}} \prp{\gpnu{}}' \prcr{\gpnu{}} \prp{\gpnu{}} \text{   .   }
\end{eqnarray*}
Here, $-\mllk{\gpdt{}|\prp{\gpsh{}}, \prp{\gpnu{}}}$ is the negative log sample GP likelihood from Section \ref{Sct2.1:GnrPrt} expressed as a function of $\gpsh{}$ and $\gpnu{}$, and $\rgh{\gpsh{}}$ and $\rgh{\gpnu{}}$ are additive roughness penalties. The values of $\lambda_{\gpsh{}}$ and $\lambda_{\gpnu{}}$ are selected using cross-validation to maximise the predictive performance of the estimated model, and bootstrap resampling is used to quantify the uncertainty of parameter estimates. The original sample is resampled with replacement a large number of times, and inference repeated for each resample. We use the empirical distributions of parameter estimates and return values over resamples as approximate uncertainty distributions.

We refer readers interested in further information on penalised likelihood methods to the work of \cite{GrnSlv94}, \cite{Dvs03}, \cite{RppEA03}, \cite{ElrMrx10}, outlined in applications to metocean by \cite{JntEwn13}.
\subsubsection*{Bayesian inference}
\begin{figure}
\begin{center}
\begin{tikzpicture}[-,>=stealth',shorten >=1pt,auto,node distance=2.5cm,
  thick,main node/.style={circle,draw,font=\sffamily\large\bfseries}, sub node/.style={circle,draw,font=\sffamily\small\bfseries}]
		\node[main node](gpdt)at(0,0){$\gpdt{}$}; 
		\node[main node](nup)at(-1,1){$\prp{\gpnu{}}$}; 
		\node[main node](xip)at(1,1){$\prp{\gpsh{}}$}; 
		\node[sub node](nupn)at(-2,2){$\prpn{\gpnu{}}$}; 
		\node[sub node](xipn)at(2,2){$\prpn{\gpsh{}}$}; 
		\node[sub node](nupna)at(-3,3){$\prpna{\gpnu{}}$}; 
		\node[sub node](nupnb)at(-2,3){$\prpnb{\gpnu{}}$}; 
		\node[sub node](xipna)at(3,3){$\prpna{\gpsh{}}$}; 
		\node[sub node](xipnb)at(2,3){$\prpnb{\gpsh{}}$}; 
		
		\path[<-,every node/.style={font=\sffamily\small}]
		(gpdt) edge node {} (nup)
					 edge node {} (xip)
		(nup)  edge node {} (nupn)
		(xip)  edge node {} (xipn)
		(nupn) edge node {} (nupna)
					 edge node {} (nupnb)
		(xipn) edge node {} (xipna)
					 edge node {} (xipnb);
\end{tikzpicture}
\end{center}
\caption{Directed acyclic graphical representation of the Bayesian inference scheme.}
\label{Fgr2}
\end{figure}
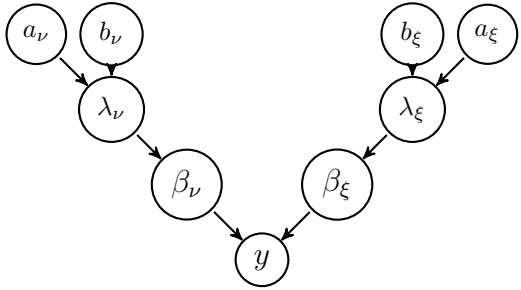

%
From a Bayesian perspective, all of $\prp{\gpsh{}}$, $\prp{\gpnu{}}$, $\prpn{\gpsh{}}$ and $\prpn{\gpnu{}}$ are treated as parameters to be estimated. Their joint posterior distribution given sample responses $\gpdt{}$ and covariates $\gpdm{}$ can be written
\begin{align*}
\pd{\prp{\gpsh{}},\prp{\gpnu{}},\prpn{\gpsh{}},\prpn{\gpnu{}}|\gpdt{},\gpdm{}} & \propto{} \pd{\gpdt{}|\gpdm{},\gpsh{},\gpsc{},\gpth{}}\pd{\prp{\gpnu{}}|\prpn{\gpnu{}}}\pd{\prp{\gpsh{}}|\prpn{\gpsh{}}}
\pd{\prpn{\gpnu{}}|\prpna{\gpnu{}},\prpnb{\gpnu{}}}\pd{\prpn{\gpsh{}}|\prpna{\gpsh{}},\prpnb{\gpsh{}}} \label{FJDist}
\end{align*}
where $\pd{\gpdt{}|\gpdm{},\gpsh{},\gpsc{},\gpth{}}$ is the sample GP likelihood from Section \ref{Sct2.1:GnrPrt} and prior distributions $\pd{\prp{\gpnu{}}|\prpn{\gpnu{}}}$, $\pd{\prp{\gpsh{}}|\prpn{\gpsh{}}}$, $\pd{\prpn{\gpnu{}}|\prpna{\gpnu{}},\prpnb{\gpnu{}}}$ and $\pd{\prpn{\gpsh{}}|\prpna{\gpsh{}}, \prpnb{\gpsh{}}}$ are specified as follows. Parameter smoothness of GP shape and (modified) scale functions is encoded by adopting Gaussian priors for their vectors  $\prp{\gpr{}}$ of basis coefficients (for $\gpr{} = \gpsh{},\gpnu{}$), expressed in terms of parameter roughness $\rgh{\gpr{}}$
\begin{equation*}
\pd{\prp{\gpr{}}|\prpn{\gpr{}}} \propto \prpn{\gpr{}}^{1/2}\exp{}\left({}-\frac{\prpn{\gpr{}}}{2}\prp{\gpr{}}\T{}\prcr{\gpr{}}\prp{\gpr{}}\right){} \text{   .   }
\end{equation*}
The roughness coefficient $\prpn{\gpr{}}$ can be seen, from a Bayesian perspective, as a parameter precision for $\prp{\gpr{}}$. It is assigned a Gamma prior
distribution, which is conjugate with the prior Gaussian distribution for $\prp{\gpr{}}$. The values of hyper-parameters are set such that Gamma priors are relatively uninformative; $a_\eta$ and $b_\eta$ takes values of $10^{-3}$ throughout this study for all parameterisations. The Bayesian inference can be illustrated by the directed acyclic graph show in Figure \ref{Fgr2}.
%
%
Estimates for $\prp{\gpsh{}}$, $\prp{\gpnu{}}$, $\prpn{\gpsh{}}$ and $\prpn{\gpnu{}}$ are obtained by sampling the posterior distribution above using MCMC. We choose to adopt a Metropolis-within-Gibbs framework (e.g. \citealt{GmrLps06}), where each of the four parameters is sampled in turn conditionally on the values of others. The full conditional distributions $\prpn{\gpsh{}}|\prp{\gpsh{}}$ and $\prpn{\gpnu{}}|\prp{\gpnu{}}$ of precision parameters are Gamma by conjugacy, and are sampled exactly in a Gibbs step. Full conditional distributions for coefficients $\prp{\gpsh{}}$ and $\prp{\gpnu{}}$ are not available in closed form; a more general Metropolis-Hastings (MH) scheme must therefore be used. \par

There are a number of potential alternative strategies regarding the MH step for $\prp{\gpr{}}$ ($\gpr{} = \gpsh{},\gpnu{}$). We choose to examine two possibilities: (a) a straightforward MH sampling of correlated Gaussian proposals for $\prp{\gpr{}}$, and (b) the mMALA algorithm of \citet{GIRO2011}, exploiting first- and second-derivative information from the log posterior to propose candidate values for the full vector of coefficients in high-probability regions. Implementations are described in the Appendix. Henceforth we refer to these two schemes as MH and mMALA respectively for brevity. The MH approach is simple to implement, but is likely to generate MCMC chains which mix relatively poorly. The mMALA scheme is expected to explore the posterior with considerably higher efficiency; however, its implementation requires knowledge of likelihood derivatives.

\subsubsection*{Comparing uncertainties}
%
Parameter uncertainty is estimated by bootstrap resampling for ML inference and by sampling from the posterior distribution of parameters for Bayesian inference. These two approaches seek to estimate parameter uncertainty, but in different ways. Since Bayesian priors are chosen to be relatively uninformative, we expect - at least naively - that inferences concerning parameter estimates and return value distributions from the two approaches will be similar, but not the same. In general, the relationship between bootstrap uncertainty estimates and those from Bayesian inference is complex, and an open topic in the literature (e.g. \citealt{FshEA05}). A thorough theoretical analysis of this relationship in the current application is beyond the scope of the current work.

\section{Evaluation of methods} \label{Sct3:EvlMth}
%
This section describes evaluation of relative performance of different model parameterisations and inference schemes introduced in Sections \ref{Sct2.2:CvrPrm} and \ref{Sct2.3:Inf}. We assess performance in terms of quality of estimation of distributions of return values corresponding to long return periods, estimated under models for large numbers of replicate samples of data from pre-specified underlying models.

We simulate 100 sample realisations, each of size 1000 from three different underlying models, described below and referred to henceforth as Cases 1, 2 and 3, and further simulate 100 sample realisations of size 5000 from the same triplet of underlying models, referring to these as Cases 4, 5 and 6 respectively. We next estimate extreme value models for all sample realisations, model parameterisations and inference schemes. We assume that any sample realisation (for any Case) corresponds to a period $\mathcal{T}$ years of observation.  We then simulate 1000 replicates of return period realisations, each replicate consisting of observations of directional extreme values corresponding to a return period of $10\times \mathcal{T}$, and estimate the distribution of the maximum observed (the $10\mathcal{T}$-year maximum return value) for all model parameterisations and inference methods, by accumulation from the 1000 replicates. Return value simulations under models estimated using Bayesian inference proceed by sampling a different vector of model parameter estimates at random from the estimated joint posterior distribution of parameters and simulating the appropriate return period of events from the corresponding distribution, for each of the 1000 replicates. For ML inference, for each replicate, we sample the vector of model parameter estimates at random from a set of 100 parameter estimate vectors of generated by the bootstrap analysis. Return value distributions are estimated omnidirectionally (that is, including all directions) and for 8 directional octants centred on the cardinal and semi-cardinal directions (by considering only those observations from the return period realisation with the appropriate directional characteristics). For each sample realisation from Cases 4, 5 and 6, we estimate return value distributions for all parameterisations but for only mMALA inference, since as will be discussed in Section \ref{Sct3.3:EffInf} below, the computational effort associated with any of mMALA, MH and MLE (with bootstrap resampling) for these Cases is large.

We quantify the quality of return value inference by comparing the empirical cumulative distribution function generated under the fitted model for each sample realisation with that from simulation under the known underlying Case. We quantify the discrepancy between empirical distribution functions by estimating Kullback-Leibler, Cramer-von Mises and Kolmogorov-Smirnov statistics. We visualise relative performance by plotting the empirical cumulative distribution function of the test statistic over the 100 sample realisations, for each combination of Case, model parameterisation and inference method. We also compare performance in terms of prediction of the 37.5$^{\text{th}}$ percentile of the $10 \mathcal{T}$-year return value distribution, since this is often used in metocean and coastal design applications; it corresponds approximately to the location of the mode of a Weibull distribution with shape parameter $\approx 2$. However, we are not only interested in quality of inference, but also in computational efficiency. This is evaluated and illustrated in Section \ref{Sct3.3:EffInf}.

To complete the full analysis described here, comprised of 100 random samples of each of 6 Cases, required running 3 dedicated workstations (exploiting each of 48 cores and 196GB RAM per workstation) for approximately 10 weeks. All assessments of return value distributions, in terms of e.g. Kullback-Leibler divergence or a central percentile, are therefore based on 100 independent estimates per Case. We note that the precision with which the median and quartiles of the distribution of Kullback-Leibler divergence, or the distribution of a central percentile (or is bias), is therefore considerably higher than that with which extreme quantiles are estimated.

In practical application to metocean design using a sample of measured or hindcast significant wave height data, for example, it is critical to demonstrate that an iterative simulation algorithm such as MH or mMALA has produced a chain which has itself converged to the stationary distribution, and whether that distribution has been adequately explored. This can be achieved, for example, by comparing inferences from multiple independent chains (e.g. \citealt{BrkGlm98}). This diagnosis is critical to such applications, and is the main evidence that MCMC inference is valid. In the current work, visual inspection of trace plots was used to confirm that an adequate period of burn-in had been specified for all combinations of parameterisation and inference scheme. However since the underlying true models are known, we choose to use comparison with the truth as the basis to assess the relative performance of different model parameterisations and inference schemes. The finding of this work is that mMALA as implemented here behaves more reasonably as a default approach to inference (requiring less user intervention and fine tuning) than MH. There is no doubt that chain convergence diagnostics would have indicated this also, and might also have prompted refinement of the MH scheme in particular to improve its performance.

\subsection{Case studies considered} \label{Sct3.1:Css}
\begin{figure}
\centering
\includegraphics[width=0.85\columnwidth]{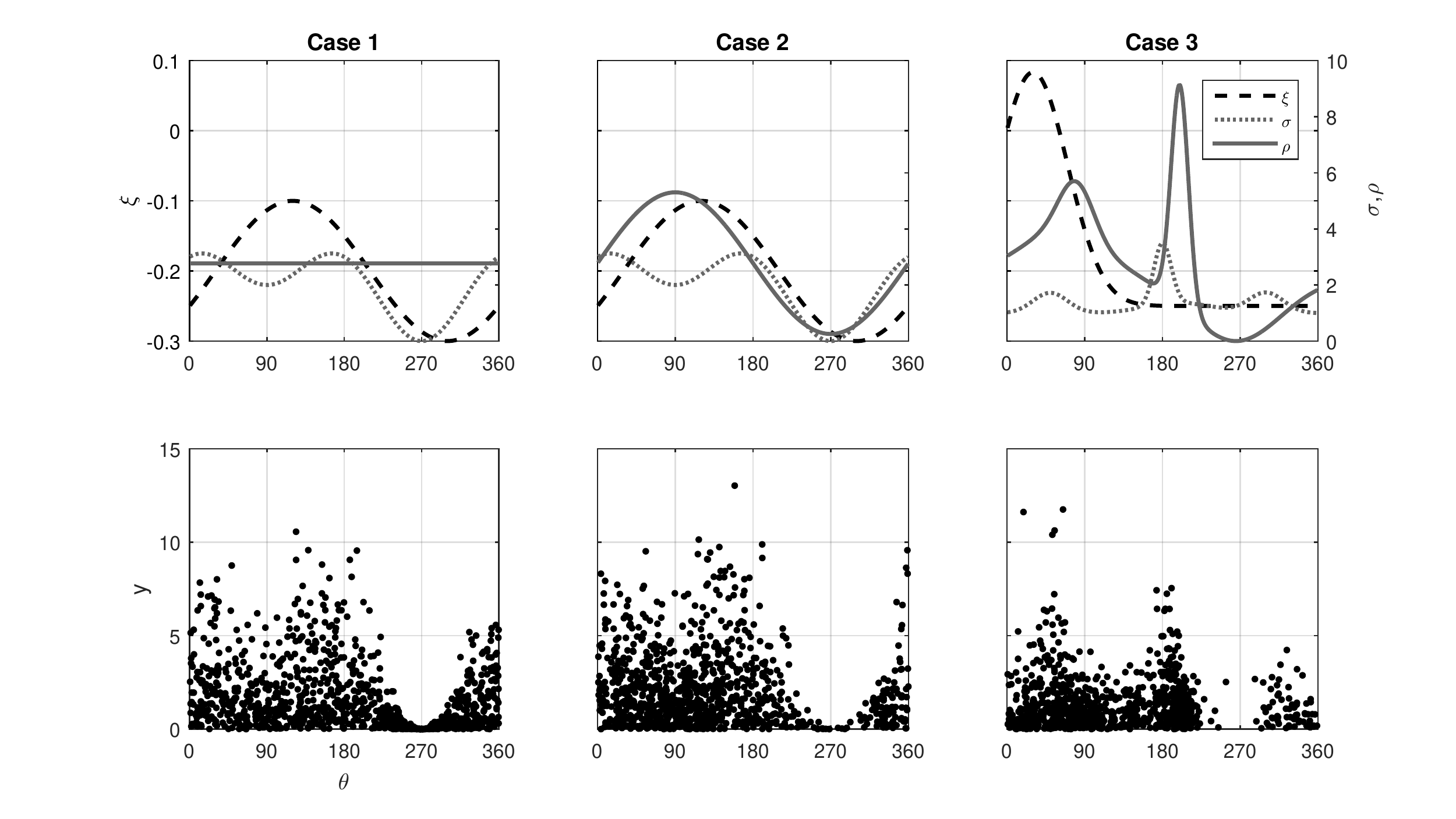}
\caption{Illustrations of sample realisations from each of Cases 1 (left), 2 (centre) and 3 (right). Upper panels show parameter variation of GP shape $\xi$, scale $\sigma$ and Poisson rate $\rho$ with direction $\theta$ for each case. Lower panels show the 10th realisation of the corresponding simulated samples. $\xi$ and $\sigma$ for Cases 4, 5 and 6 are identical to those of Cases 1, 2 and 3 respectively. The value of $\rho$ for Cases 4, 5 and 6 is five times that of Cases 1,2 and 3 respectively. \label{Fgr3}}
\end{figure}

%
First, we describe model Cases 1-6 used to generate sample realisations for extreme value modelling. For each Case, potentially all of Poisson rate $\gprt{}$ of threshold exceedance, GP shape $\gpsh{}$ and scale $\gpsc{}$ of exceedance size vary as a function of covariate $\gpcv{}$. The extreme value threshold $\gpth{}$ is fixed at zero throughout.
\paragraph{Case 1}: For extreme value threshold $\gpthf{\gpcv{}} = 0$, we simulate $1000$ observations with a uniform Poisson rate $\gprtf{\gpcv{}} = 1000/360$ per degree covariate, and a low order Fourier parameterisation of GP shape $\gpshf{\gpcv{}} = \sin{}(\gpcv{}) + \cos{}(2\gpcv{}) + 2$ and scale $\gpscf{\gpcv{}} = - 0.2 + (\sin{}(\gpcv{}-30))/10 $.
\paragraph{Case 2}: For extreme value threshold $\gpthf{\gpcv{}} = 0$ and the same Fourier parameterisation of GP shape and scale as in Case 1, a non-uniform Poisson rate $\gprtf{\gpcv{}} = \max{}\left({}\sin{}(\gpcv{})+1.1,0\right) \times 1000/c_\rho$, where $c_\rho=\int_0^{360} \max{}\left({}\sin{}(\gpcv{})+1.1,0\right){} d\gpcv{}$ is used to simulate $1000$ observations.
\paragraph{Case 3}: For extreme value threshold $\gpthf{\gpcv{}} = 0$, the forms of each of $\gprtf{\gpcv{}}$, $\gpshf{\gpcv{}}$ and $\gpscf{\gpcv{}}$ are defined by mixtures of between one and five Gaussian densities, as illustrated in Figure \ref{Fgr3}. Sample size is $1000$.
\paragraph{Cases 4, 5 and 6}: These cases are identical to Cases 1, 2 and 3 respectively, except that Poisson rate $\rho$ is increased by a factor of five. Sample size is therefore $5000$.

\begin{figure}
\centering
\includegraphics[width=0.85\columnwidth]{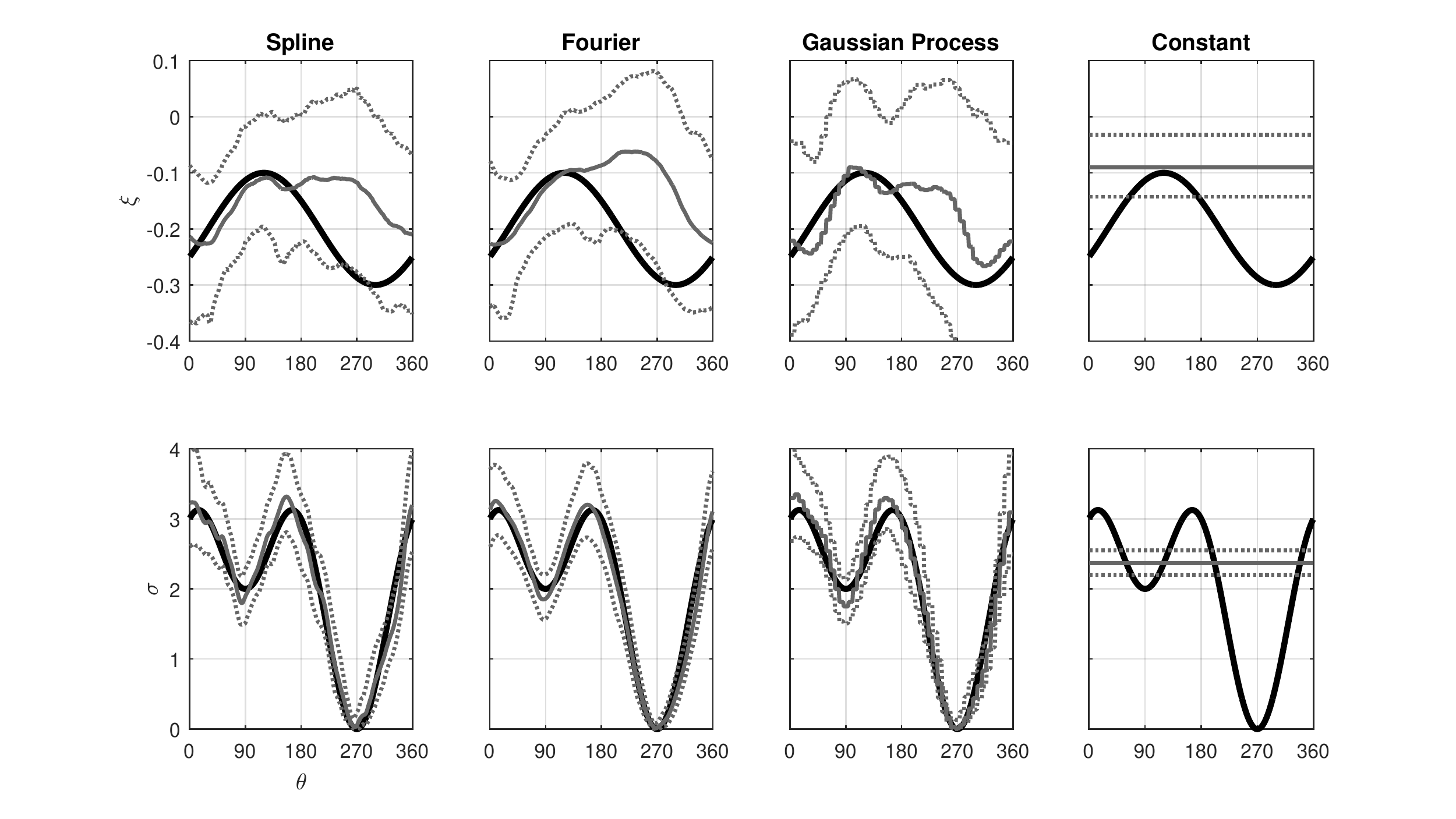}
\caption{Parameter estimates for GP shape $\xi$ (upper) and scale $\sigma$ (lower) for the sample realisation of Case 2 shown in Figure \ref{Fgr1}, for different model parameterisations (left to right: Spline, Fourier, Gaussian Process, Constant) using mMALA inference. Each panel illustrates the true parameter (solid back), posterior median estimate (solid grey) with 95\% credible interval (dashed grey). \label{Fgr4}}

\end{figure}

\begin{figure}
\centering
\includegraphics[width=0.85\columnwidth]{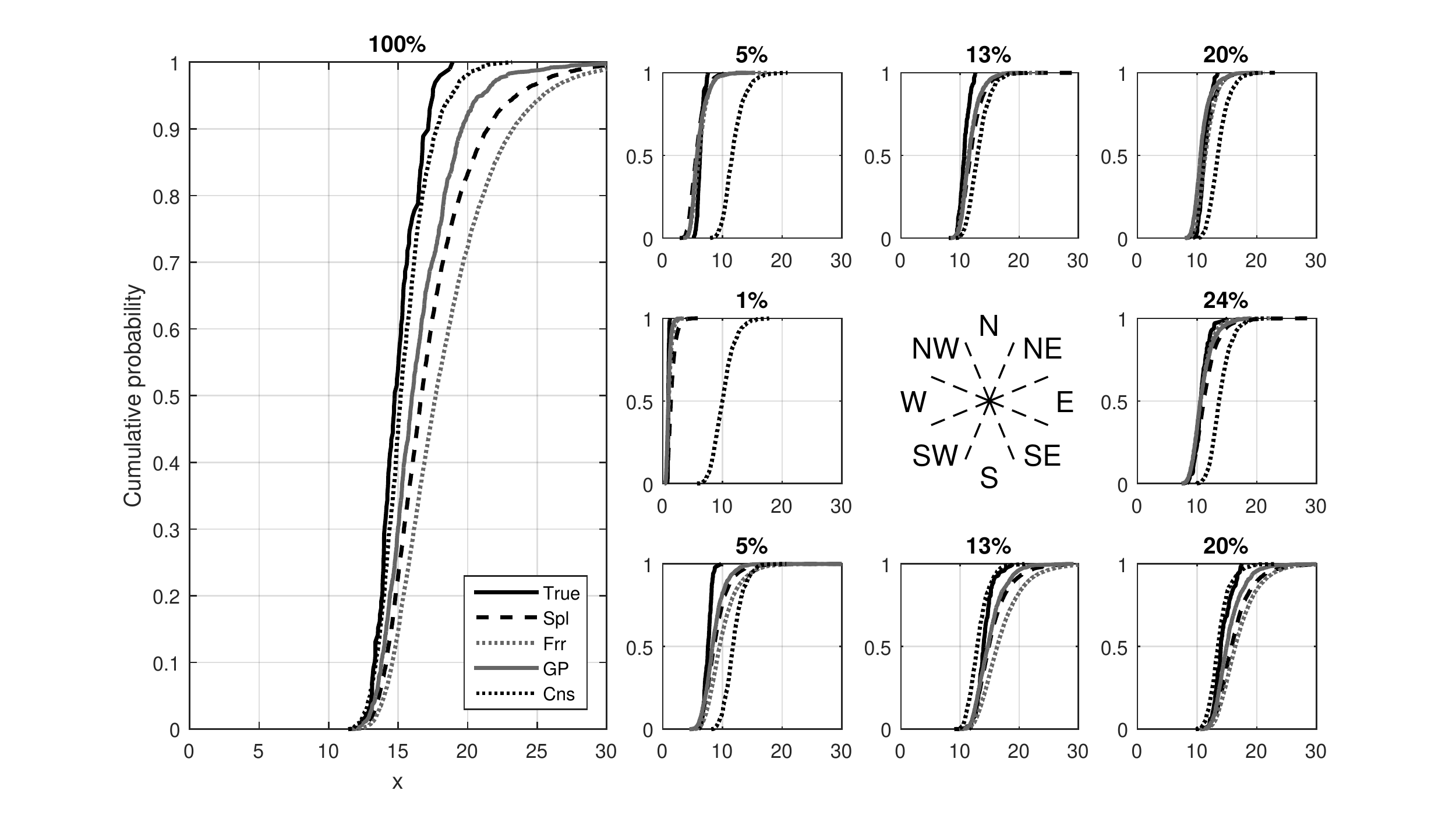}
\caption{Posterior cumulative distribution functions of return value for the sample realisation of Case 2 shown in Figure \ref{Fgr1}, corresponding to a return period of ten times the period of the original sample. The left hand panel shows the omnidirectional return value distribution, and right hand panels the corresponding directional estimates. The title for each panel gives the expected percentage of individuals in that directional sector. In each panel, estimates are given for different model parameterisations using mMALA inference. The true return value distribution is given in solid black.}
\label{Fgr5}
\end{figure}

Figure \ref{Fgr3} illustrates typical sample realisations of Cases 1, 2 and 3. Parameter variation of GP shape $\xi$ and scale $\sigma$ with direction $\theta$ are identical in Cases 1 and 2. Poisson rate $\rho$ is constant in Case 1 only. In Cases 2 and 3, $\rho$ is very small at $\theta \approx 270^\circ$ leading to a sparsity of corresponding observations. $\xi$ is largest (but negative) at $\theta \approx 120^\circ$ for Cases 1 and 2, leading to larger observations here. For Case 3, $\xi$ is largest (and positive) at $\theta \approx 30^\circ$ leading to the heaviest tail in any of the Cases considered. Figure \ref{Fgr4} shows parameter estimates for $\xi$ and $\sigma$, corresponding to the sample realisation of Case 2 shown in Figure \ref{Fgr3}, for different model parameterisations using mMALA inference. Visual inspection suggests that estimates of similar quality are obtained using all of Spline, Fourier and Gaussian Process parameterisations, but that the Constant parameterisation is poor. It is also apparent that identification of $\xi$ is more difficult than $\sigma$. Corresponding plots (not shown) for  maximum likelihood and Metropolis-Hastings inference show broadly similar characteristics, as do plots for other realisations of the same Case, and realisations of different Cases. (Posterior) cumulative distribution functions of return values based on models for the sample realisations of Case 2 illustrated in Figures \ref{Fgr3} and \ref{Fgr4}, corresponding to a return period of ten times the period of the original sample, are shown in Figure \ref{Fgr5} for different model parameterisations and mMALA inference.

\begin{figure}
\centering
\includegraphics[width=0.85\columnwidth]{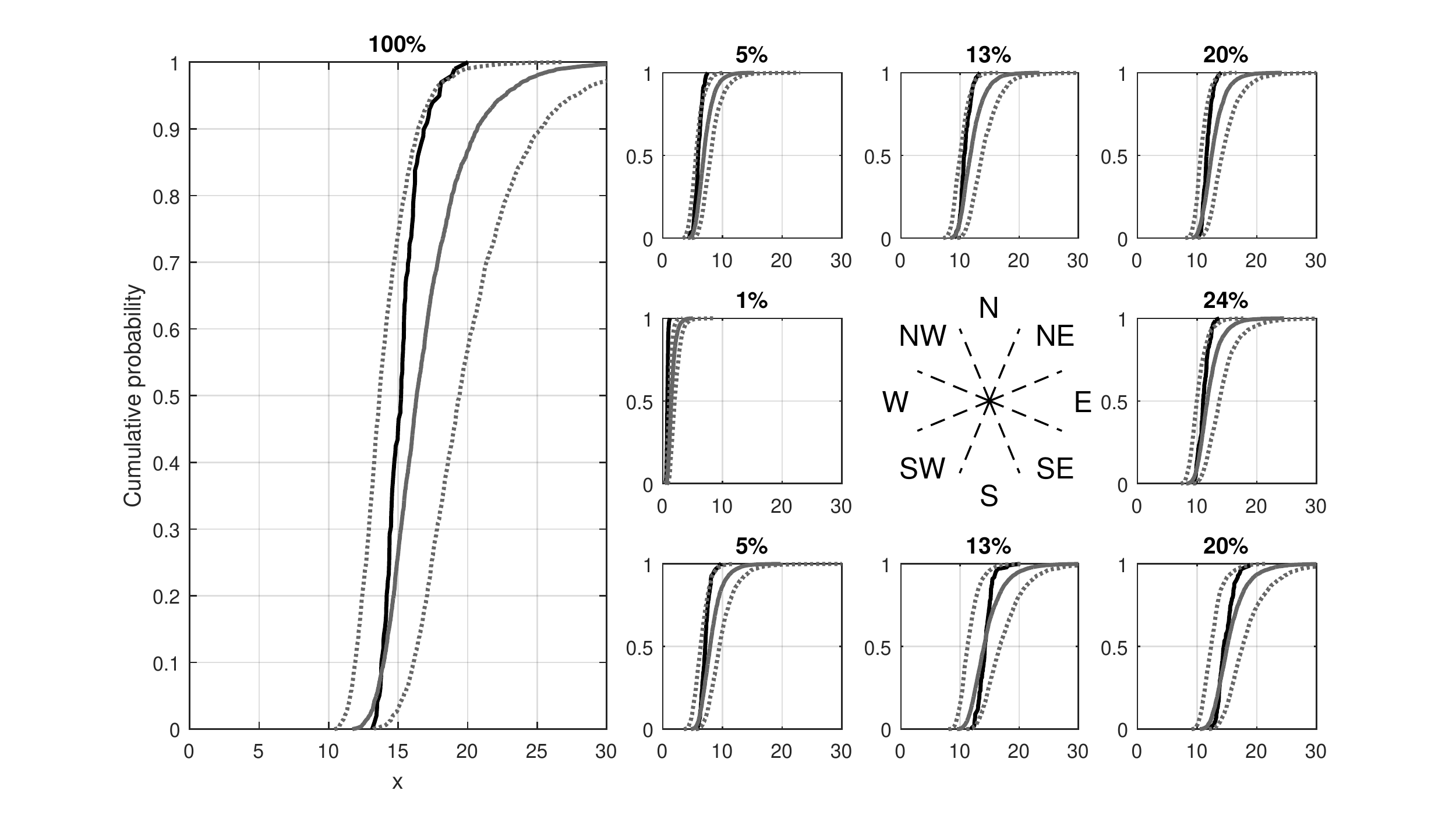}
\caption{Uncertainty (over all 100 sample realisations) in the cumulative distribution function of return value for Case 2, corresponding to a return period of ten times the period of the original sample. The left hand panel shows the omnidirectional return value distribution, and right hand panels the corresponding directional estimates. The title for each panel gives the expected percentage of individuals in that directional sector. In each panel, the true return value distribution is given in solid black. The median estimate (over realisations) for return value distribution of the Spline model parameterisation using mMALA inference is shown in solid grey, with corresponding point-wise 95\% uncertainty band in dashed grey.}
\label{Fgr6}
\end{figure}

It can be seen omnidirectionally that the Constant parameterisation provides best agreement with the known return value distribution, despite the fact that parameter estimates in Figure \ref{Fgr4} do not reflect the directional non-stationarity present. In \cite{JntEwnFrr08a}, it is demonstrated that a stationary extreme value model may produce good estimates for omni-directional return value distributions in at least two situations. Firstly, when the extreme value threshold is set sufficiently high that only observations from the most extreme interval of the covariate domain are modelled, therefore threshold exceedances and estimated models will be effectively stationary. Secondly, it may be that different bias effects introduced by the stationarity assumption compensate for each other in estimation of return values at a certain return levels. For example, with reference to the sample illustrated in Figure \ref{Fgr4}, the constant model underestimates the value of GP shape in general, but does a relatively good job of identifying the maximum values of shape and scale in directions around $120^\circ$. The constant model does a very poor job for directions around $270^\circ$. The uncertainty in parameter estimates from the constant model is also lower since the number of parameters to estimate is lower for that model. Reference to Figure \ref{Fgr3} for Case 2 shows also that the rate of occurrence of threshold exceedances is larger for directions where the constant model performs well. These characteristics are reflected in the corresponding return value estimates in Figure \ref{Fgr5}. For directional octants, the constant model does relatively well for directions around $120^\circ$, but very poorly around $270^\circ$. However, the omnidirectional estimate is dominated by directions around $120^\circ$. The relatively good performance of the directional model around $120^\circ$, together with the large rate of occurrence of events there and the relatively small parameter uncertainties for the constant model, result in its providing good estimation (in this case) of the omnidirectional return value. This occurs despite the fact that the directional bias of the constant model is greater than that of any of the non-stationary models. In general however, it is not possible to know a priori how a constant model will perform in estimating the omnidirectional value. It is clear however that its directional bias will be larger than that of an appropriate directional model. Reference to Figure \ref{Fgr9} below, assuming that mMALA provides more satisfactory inference as implemented here, suggests that values of KL divergence are somewhat larger and more variable for the constant model compared to the Spline or Gaussian process parameterisations. Reference to Figure \ref{Fgr10} shows that the Constant model does very poorly for certain directional sectors. Omnidirectionally, and for 8 directional octant sectors, non-stationary model parameterisations perform similarly. However, it is clear that the Constant parameterisation does particularly poorly for the western and north-western sectors, for which the rate of occurrence of events is relatively low, and both $\xi$ and $\sigma$ are near their minimum values. Figure \ref{Fgr6} illustrates uncertainty (over all 100 sample realisations) in the cumulative distribution function of return value for Case 2, corresponding to a return period of ten times the period of the original sample, using the Spline parameterisation and mMALA inference. The median estimate for the return value distribution (over all 100 sample realisations of Case 2) is shown in solid grey, with corresponding point-wise 95\% uncertainty band in dashed grey. The true return value distribution is given in solid black. There is good agreement in all sectors. We explore differences in inferences for return value distributions more fully in Section \ref{Sct3.2:QltInf}.
%
%
%
%

\subsection{Assessing quality of inference} \label{Sct3.2:QltInf}
%
The criteria used to compare distributions of return values are now described. Since, for comparison only, we only have access to samples from distributions, where necessary we project empirical distributions onto a linear grid using linear interpolation, and evaluate grid-based approximations to facilitate comparison. Then we compare empirical return value distributions using each of the following three statistics. The Kolmogorov-Smirnov criterion compares two distributions in terms of the maximum vertical distance between cumulative distribution functions, as $\Dks{\DFna{0},\DFna{1}} = \sup_{x}|\DF{1}{x}-\DF{0}{x}|$ . The Cramer-von Mises criterion evaluates the average squared difference of one distribution from a second, reference distribution, using $\Dcvm{\DFna{0},\DFna{1}} = \int_{-\infty{}}^{\infty{}} \left({}\DF{1}{x}-\DF{0}{x}\right)^{2}\Df{0}{x} dx$. The Kullback-Leibler divergence compares distributions using the average ratio of logarithms of density functions $ \Dkl{\DFna{0},\DFna{1}} = \int_{-\infty{}}^{\infty{}} \log{}\left({}\frac{\Df{0}{x}}{\Df{1}{x}}\right){}\Df{0}{x} dx$; in this work, we use the approximation of \citet{PERE2008}. The general characteristics of differences in return value inference due to model parameterisation and inference method were found to be similar for each of the three statistics. Only comparisons using Kullback-Leibler (KL) divergence are therefore reported here. We note that perfect agreement between $\Df{1}{x}$ and $\Df{0}{x}$ yields a minimum KL divergence of zero.\par

\begin{figure}
\centering
\includegraphics[width=0.85\columnwidth]{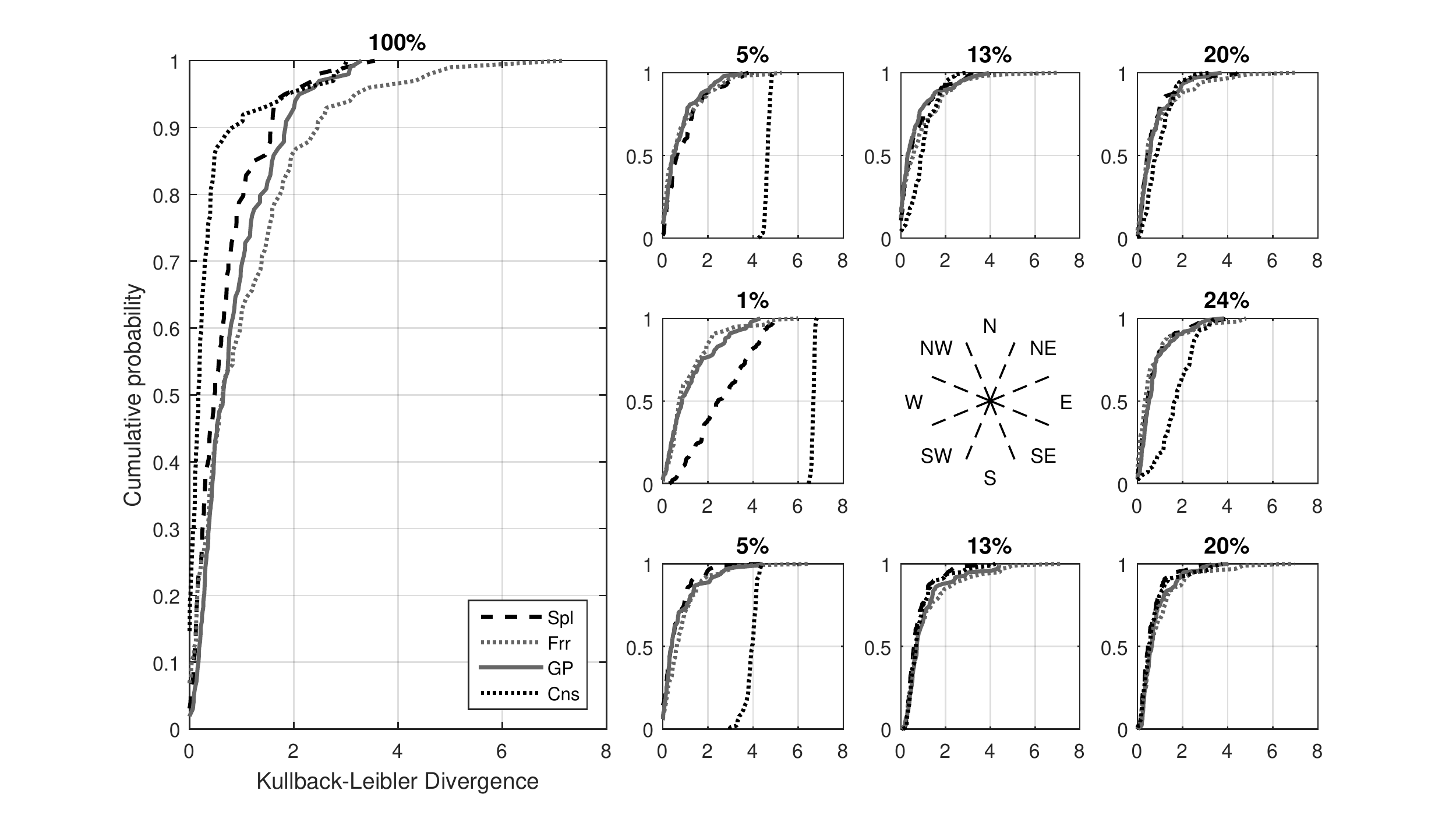}
\caption{Empirical cumulative distribution functions of the Kullback-Leibler divergence between return value distributions (corresponding to a return period of ten times that the original sample) estimated under the true model and those estimated under models  of sample realisations with different parameterisations and mMALA inference for Case 2. The title for each panel gives the expected percentage of individuals in that directional sector.}
\label{Fgr7}
\end{figure}

\begin{figure}
\centering
\includegraphics[width=0.85\columnwidth]{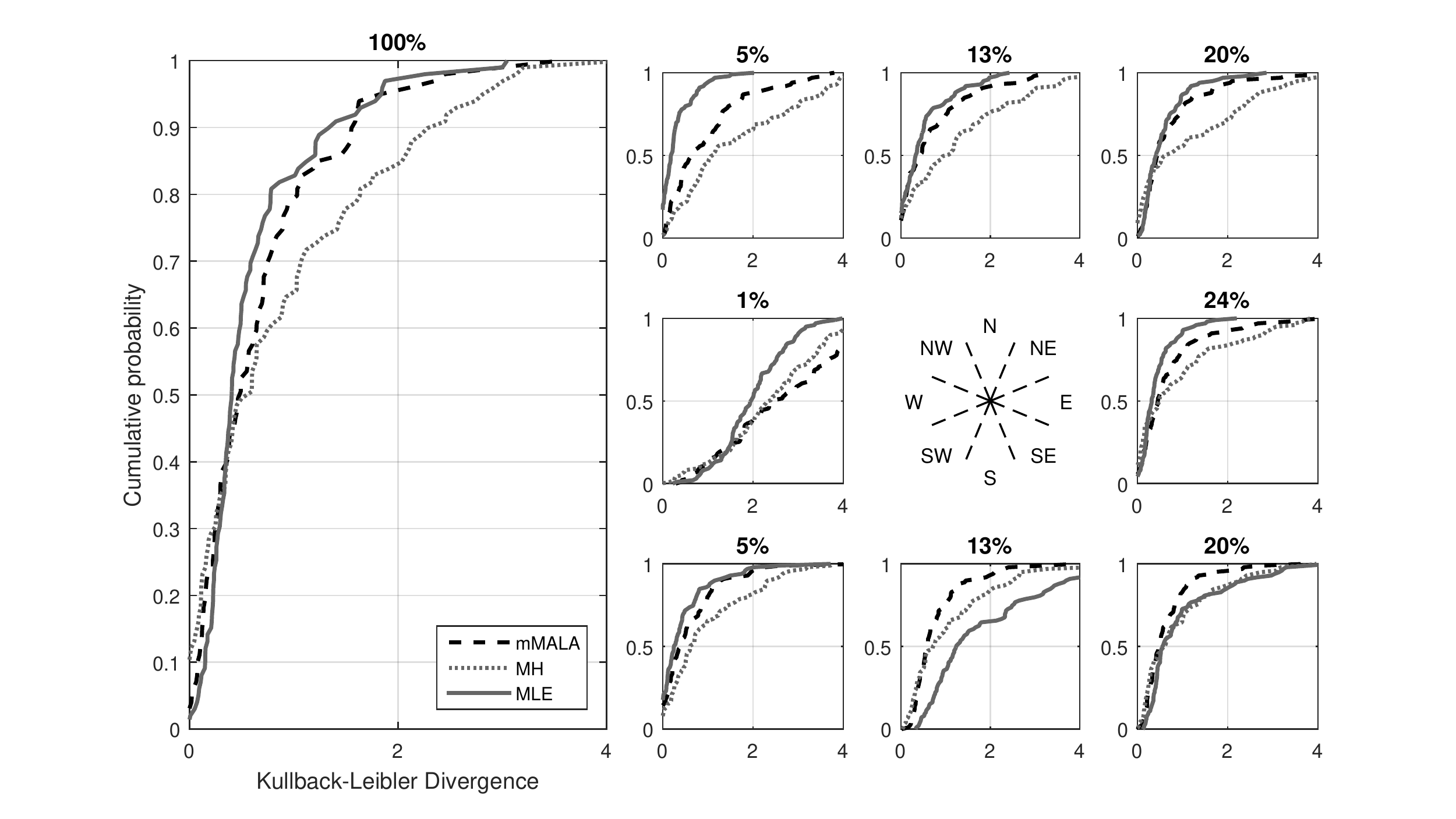}
\caption{Empirical cumulative distribution functions of the Kullback-Leibler divergence between return value distributions (corresponding to a return period of ten times that the original sample) estimated under the samples from the true return value distribution and those estimated under models  of sample realisations with Spline parameterisations and different inference procedures for Case 2. The title for each panel gives the expected percentage of individuals in that directional sector.}
\label{Fgr8}
\end{figure}

For illustration, Figure \ref{Fgr7} shows empirical cumulative distribution functions for the KL divergence between return value distributions (corresponding to a return period of ten times that the original sample) estimated under the true model and those estimated under models  of sample realisations with different parameterisations and mMALA inference for Case 2. The distributions of KL divergence from all non-stationary model parameterisations appear to be very similar, as might be expected from consideration of figures similar to Figure \ref{Fgr5}. However, the Constant model yields the best performance omnidirectionally in this case (since the corresponding distribution of KL divergence is shifted towards zero). In stark contrast, the Constant model does particularly badly in the eastern, south-western, western and north-western sectors. Figure \ref{Fgr8} gives empirical cumulative distribution functions of the KL divergence between return value distributions (corresponding to a return period of ten times that of the original sample) estimated under the true return value distribution and those estimated under models of sample realisations with Spline parameterisations and different inference procedures for the same Case. There appears to be little to choose between mMALA and MLE inference methods for this Case, with MH somewhat poorer. \par

\begin{figure}
\centering
\includegraphics[width=0.85\columnwidth]{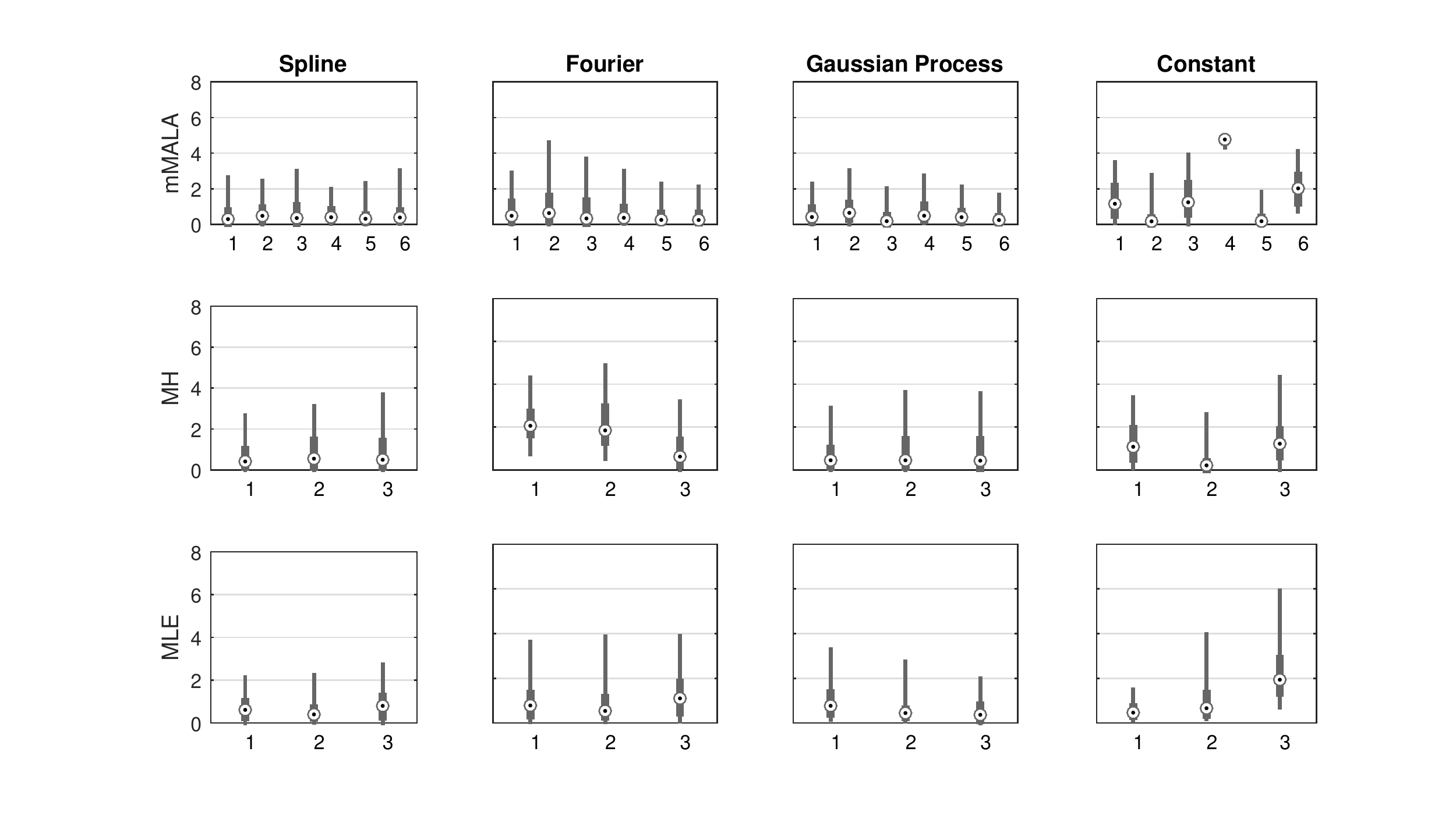}
\caption{Box-whisker comparison of samples of Kullback-Leibler (KL) divergence between omnidirectional return value distributions (corresponding to a return period of ten times that the original sample) estimated under samples from the true return value distribution and those estimated under models of each of 100 sample realisations. mMALA inference is reported for all six Cases (abscissa labels) and model parameterisations (columns of panels). Metropolis-Hastings (MH) inference and maximum likelihood estimation (MLE) are reported only for the smaller sample sizes. The sample of KL divergence is summarised by the median (white disc with black central dot), the interquartile range (grey rectangular box) and the (2.5\%, 97.5\%) interval (grey line).}
\label{Fgr9}
\end{figure}

\begin{figure}
\centering
\includegraphics[width=0.85\columnwidth]{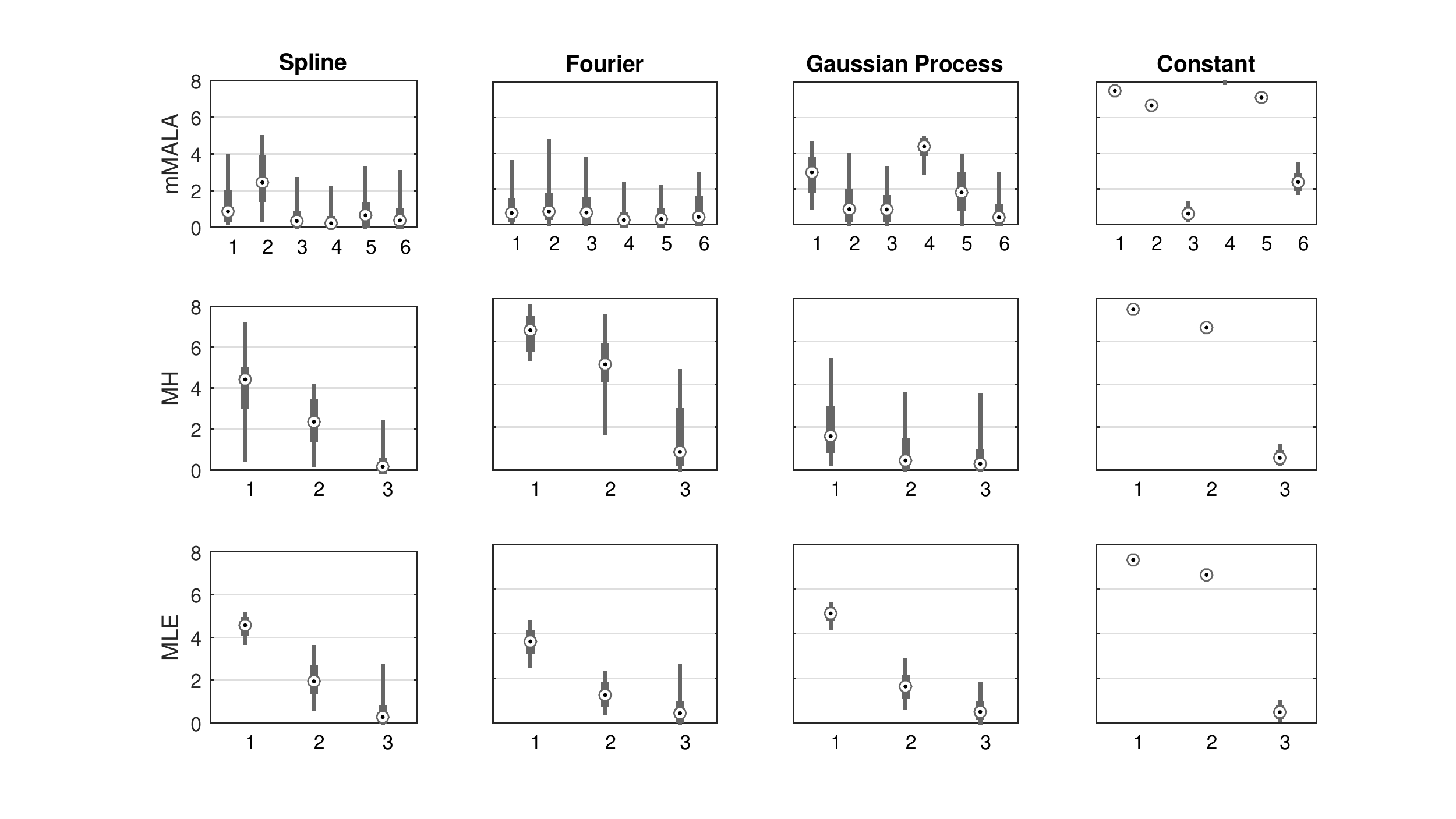}
\caption{Box-whisker comparison of samples of Kullback-Leibler (KL) divergence between western sector return value distributions (corresponding to a return period of ten times that the original sample) estimated under samples from the true return value distribution and those estimated under models of each of 100 sample realisations. mMALA inference is reported for all six Cases (abscissa labels) and model parameterisations (columns of panels). Metropolis-Hastings (MH) inference and maximum likelihood estimation (MLE) are reported only for the smaller sample sizes. The sample of KL divergence is summarised by the median (white disc with black central dot), the interquartile range (grey rectangular box) and the (2.5\%, 97.5\%) interval (grey line). The ordinate scale is the same as that of Figure \ref{Fgr9} to facilitate comparison. The Constant parameterisation for Case 4 with mMALA inference yields values of KL divergence larger than 8.}
\label{Fgr10}
\end{figure}

Figure \ref{Fgr9} summarises the characteristics of distributions for KL divergence corresponding to the omnidirectional return value distribution for all Cases, model parameterisations and inference methods considered in this work. In general, we note that all non-stationary parameterisations perform well with mMALA inference. With MH inference, performance is generally poorer, especially for Fourier parameterisation. MLE does better than MH. We note that the Constant parameterisation generally performs well for the omnidirectional return value, but there is some erratic behaviour, notably for Case 4. Figure \ref{Fgr10} is the corresponding plot for the (generally sparsely populated) western directional sector. The Spline and Fourier parameterisations with mMALA inference perform best. We note that the Fourier parameterisation does less well using MH and MLE, and that the Constant parameterisation behaves very erratically. We also note that, somewhat surprisingly, the Gaussian Process model performs considerably less well than the Spline and Fourier parameterisations. We surmise that this is due to mean-reversion in the absence of observations, compared with the other parameterisations which prefer interpolation to reduce parameter roughness. We note that the Constant parameterisation also performs badly for Case 4.

%
%
%

From a practitioner's perspective, it is also interesting to quantify the performance of different model parameterisations and inference methods in estimating some central value (e.g. the mean, median, of 37.5$^{\text{th}}$ percentile) of the distribution of the return value. We choose the 37.5$^{\text{th}}$ percentile, since this percentile is near the mode of the distribution, and commonly used in the met-ocean community.

\begin{figure}
\centering
\includegraphics[width=0.85\columnwidth]{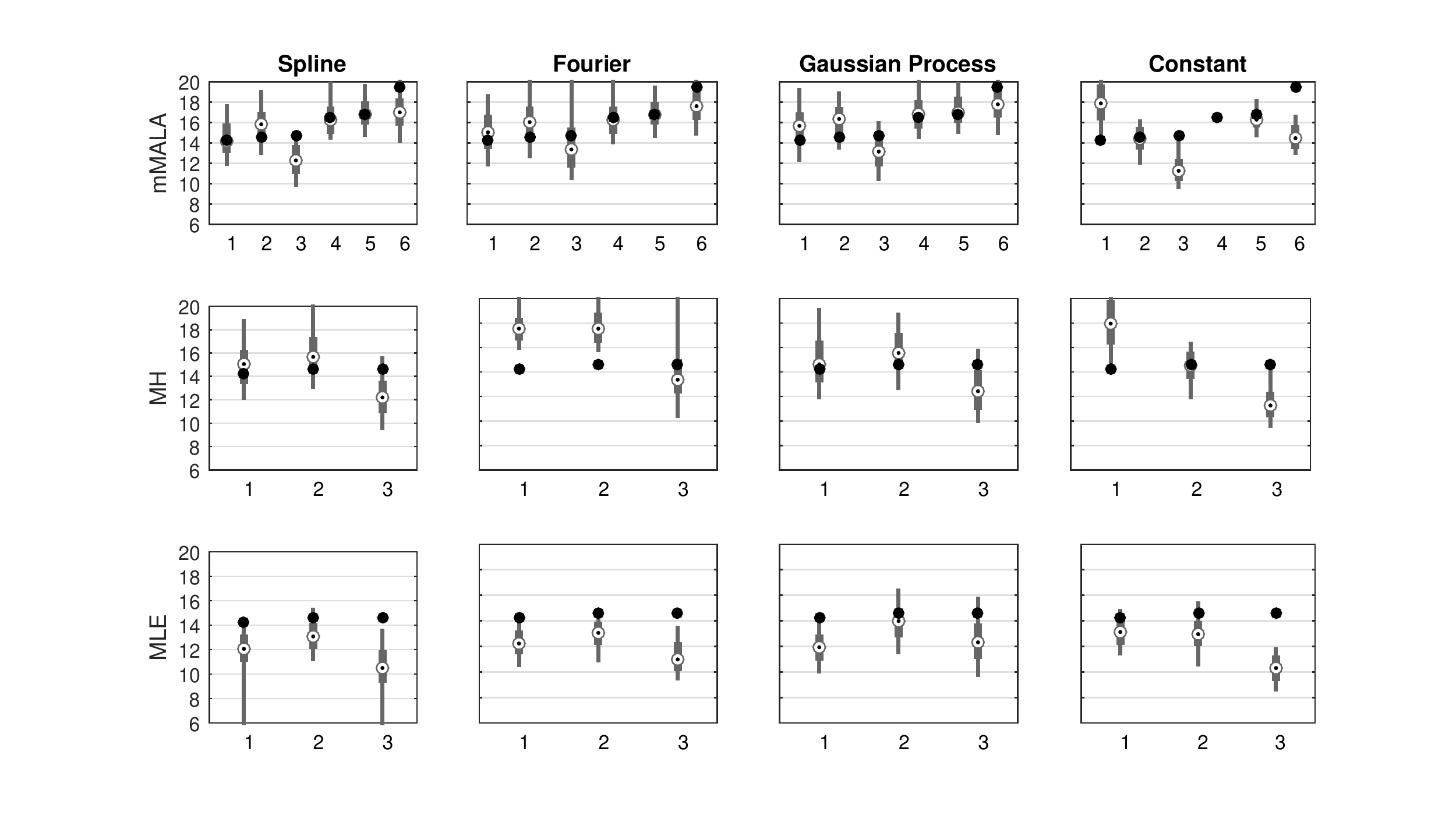}
\caption{Box-whisker comparison of estimates for the 37.5$^{\text{th}}$ percentile of the omnidirectional return value distribution (in metres) for different cases, model parameterisations and inference procedures. mMALA inference is reported for all six cases (abscissa labels) and model parameterisations (columns of panels). Metropolis-Hastings (MH) inference and maximum likelihood estimation (MLE) are reported only for the smaller sample sizes. In each panel, the estimate from simulation under the true model is shown as a black disc. The distribution of estimates from 100 different sample realisations is summarised by the median value (white disc with black central dot), the interquartile range (grey rectangular box) and the (2.5\%, 97.5\%) interval (grey line). The Constant parameterisation for Case 4 with mMALA inference yields values larger than 20m.}
\label{Fgr11}
\end{figure}

Figure \ref{Fgr11} illustrates this comparison in terms of a box-whisker plot. In each panel, the true value, estimated by simulation under the true model, is shown as a black disc. The distribution of estimates from 100 different sample realisations of each Case is summarised by the median (white disc with black central dot), the interquartile range (grey rectangular box) and the (2.5\%, 97.5\%) interval (grey line).

\begin{figure}
\centering
\includegraphics[width=0.85\columnwidth]{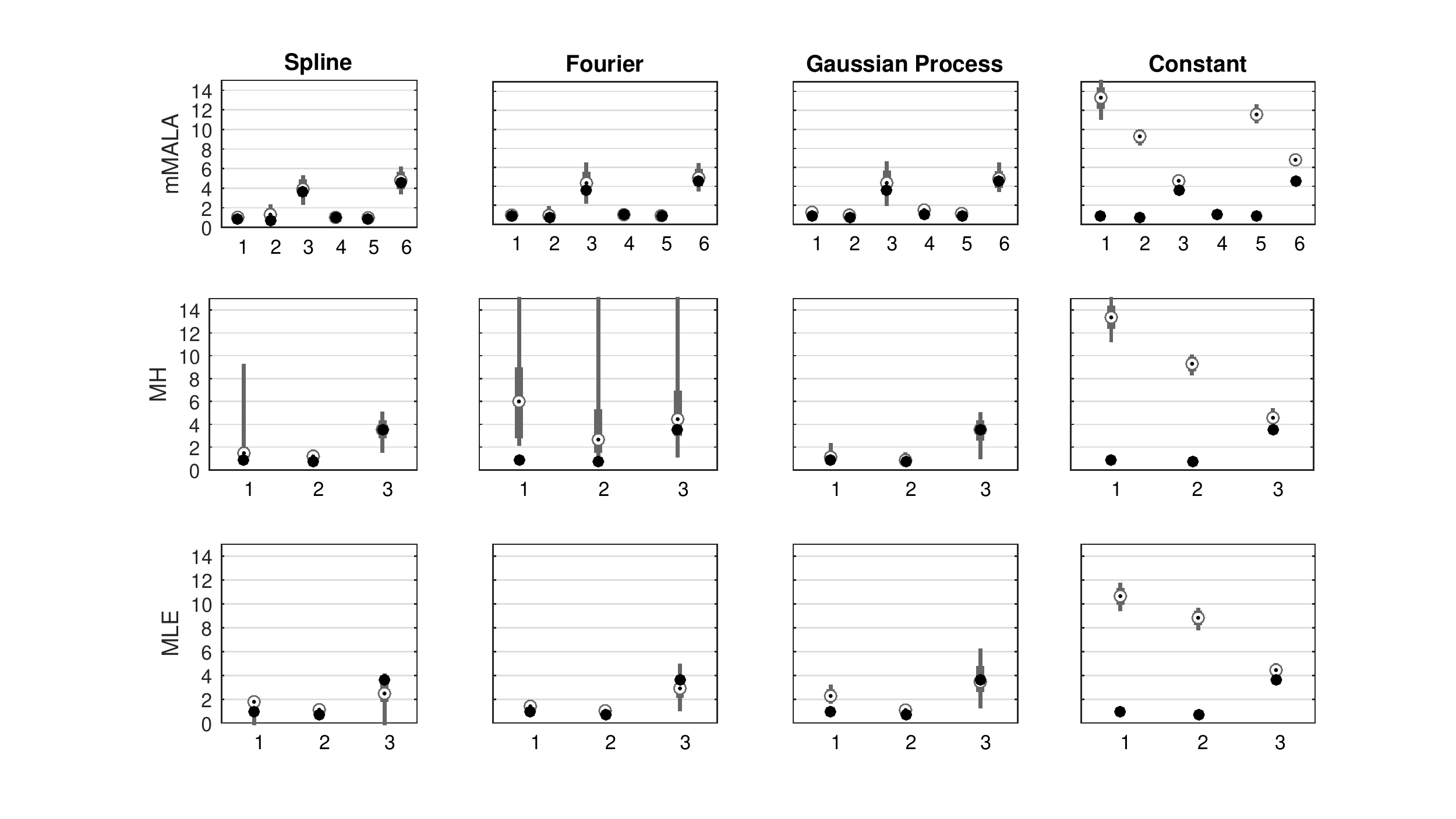}
\caption{Box-whisker comparison of estimates for the 37.5$^{\text{th}}$ percentile of the return value distribution (in metres) for the western directional sector (least populous for cases 2, 3, 5 and 6), for different cases, model parameterisations and inference procedures. mMALA inference is reported for all six cases (abscissa labels) and model parameterisations (columns of panels). Metropolis-Hastings (MH) inference and maximum likelihood estimation (MLE) are reported only for the smaller sample sizes. In each panel, the estimate from simulation under the true model is shown as a black disc. The distribution of estimates from 100 different sample realisations is summarised by the median value (white disc with black central dot), the interquartile range (grey rectangular box) and the (2.5\%, 97.5\%) interval (grey line). The Constant parameterisation for Case 4 with mMALA inference yields values larger than 15m.}
\label{Fgr12}
\end{figure}

The performance of non-stationary parameterisations with mMALA is good, with the possible exception of Case 3 (and Case 6); yet MH inference tends to produce greater bias and variability in estimates of the 37.5$^{\text{th}}$ percentile. We may surmise that this difference may be due to the fact that mMALA exploits knowledge of likelihood gradient and curvature. The Constant model again performs more erratically. Corresponding box-whisker plots for the eastern directional octant (not shown) show similar characteristics: for a given inference scheme, all non-stationary models yield similar performance, but the Constant model overestimates throughout. It is interesting that MLE shows some bias in return value estimation for the omnidirectional 37.5$^{\text{th}}$ percentile, but this is not the case in general. True values of the 37.5$^{\text{th}}$ percentile for the western sector (see Figure \ref{Fgr12}) are considerably lower than for the eastern sector, and lower again than the omnidirectional values. In this sector, the rate of occurrences of events is generally lower in all cases. Nevertheless, Figure \ref{Fgr12} has many similar features to Figure \ref{Fgr11}. However, we note that MH struggles in combination with the Fourier parameterisation, probably since the latter has the whole of the covariate domain as its support; intelligent proposals (like those used here in MLE and mMALA) are necessary. Estimates using the Constant model are erratic. Overall, we note that MLE and mMALA inference for all of Spline, Fourier and Gaussian Process parameterisations perform relatively well, and equally well.
%
%

\subsection{Assessing efficiency of inference} \label{Sct3.3:EffInf}
\begin{figure}
\centering
\includegraphics[width=1.0\columnwidth]{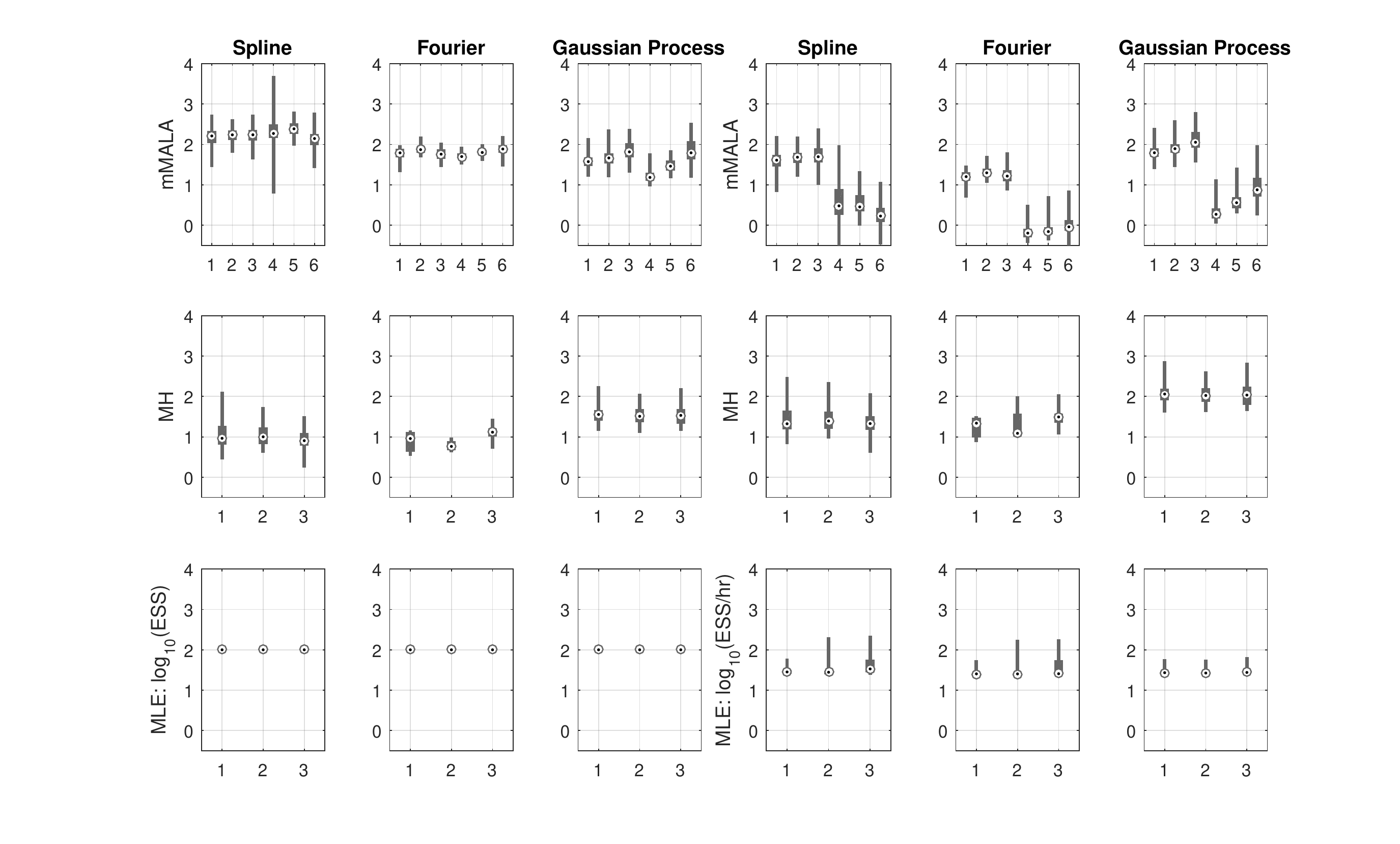}
\caption{Estimates for effective sample size (ESS, left hand side) and effective sample size per hour (ESS/hr, right hand side) on logarithm base 10 scale for different cases, non-stationary model parameterisations and inference procedures. mMALA inference is reported for all six cases (abscissa labels) and model parameterisations (columns of panels). Metropolis-Hastings (MH) inference and maximum likelihood estimation (MLE) are reported only for the smaller sample sizes. The distribution of estimates from 100 different realisations of the original sample is summarised by the median value (white disc with black central dot), the interquartile range (grey rectangular box) and the (2.5\%, 97.5\%) interval (grey line).}
\label{Fgr13}
\end{figure}

%
The effective sample size ($m^*$, e.g. \citealt{Gyr92}) gives an estimate of the equivalent number of independent iterations that a Markov chain Monte Carlo represents, and is defined by $m^* = m/(1+2\sum_{k=1}^\infty c_k)$, where $c_k$ is the autocorrelation of the MCMC chain at lag $k$, and $m$ is the actual chain length. The effective sample size per hour is defined by $m^*/T$, where $T$ is the elapsed computational time (in hours) for $m$ steps of the chain. For maximum likelihood inference with bootstrap uncertainty estimation, since bootstrap resamples are independent of one another, we estimate effective sample size per hour as $m_{BS}/T$ where $m_{BS}$ is the number of bootstrap resamples used and $T$ is now the total elapsed computational time (in hours) to execute analysis of the $m_{BS}$ bootstrap resamples. Comparison of effective sample sizes per hour for different cases, parameterisations and inference methods gives some indication of relative computational efficiency, although objective comparison is difficult. In particular we note that software implementations in MATLAB exploiting common computational structures between different approaches have been used; these are almost certainly to the detriment of computational efficiency for some of the approaches, particulary the Constant parameterisation. For this reason, we do not report effective sample size per hour for the Constant parameterisation. Computational run-times are also of course critically dependent on software and hardware resources used. We note that the focus of this work is primarily quality of inference, rather than its computational efficiency. Specific modelling choices, such as the set-up of the cross-validation strategy adopted for MLE and choice of burn-in length and proposal step-size for MH and mMALA within reasonable bounds may not influence inferences greatly, but will obviously however affect run times. Similarly the Spline, Fourier and Gaussian Process parameterisations used were chosen to be of similar complexity, but small differences may again influence relative computational efficiency of inference. With these caveats in mind, Figure \ref{Fgr13} illustrates the distribution of estimated effective sample size (ESS), and effective sample size per hour (ESS/hr) for different cases, model parameterisations and inference procedures.

The left hand side of Figure 12 illustrates $\log_{10}(ESS)$ for all combinations of parameterisations and inference schemes. For MLE inference, we choose to report the number of bootstrap resamples used, as described in Section 3. The effective sample sizes for mMALA inference are considerably larger than for MH for B-spline and Fourier parameterisations. For the Gaussian process parameterisation, mMALA still provides a larger ESS, but the difference between mMALA and MH is smaller. For mMALA inference, ESS is largest for the B-spline parameterisation; the Gaussian process parameterisation provides the smallest ESS on average. For MH inference, ESS for B-splines and Fourier parameterisations is near 10, suggesting that the posterior density has not been sufficiently explored due to poor MCMC mixing using the MH algorithm as implemented. The value of ESS is approximately constant across the different Cases examined for all combinations of model parameterisation and inference method. The right hand side of Figure 12 indicates that for B-spline parameterisation, ESS/hr is also higher for mMALA than for MH. For Fourier and Gaussian process parameterisations, ESS/hr is comparable for mMALA and MH. The right hand side of Figure \ref{Fgr13} shows that, for mMALA inference, the ESS/hr is considerably lower for Cases 4, 5 and 6, indicating that inference using large sample sizes is slower. For this reason, in this work, we do not provide results for Cases 4, 5 and 6 using MLE and MH. Overall, comparing non-stationary parameterisations, ESS/hr is larger for Splines and Gaussian Processes than for Fourier. There is little difference in ESS/hr for different model parameterisations using MLE.
%

\section{Discussion}\label{Sct4:Dsc}
%
Adequate allowance for non-stationarity is essential for realistic environmental extreme value inference. Ignoring the effects of covariates leads to unrealistic inference in general. The applied statistics and environmental literature provides various competing approaches to modelling non-stationarity. Adoption of non- or semi-parametric functional forms for generalised Pareto shape and scale in peaks over threshold modelling is far preferable in general in real world applications, than the assumption of a less flexible parameterisation. We find that B-spline and Gaussian Process parameterisations estimated by Bayesian inference (using mMALA) perform well in terms of quality and computational efficiency, and generally outperform alternatives in this work.\par

The Gaussian Process parameterisation is computationally unwieldy for larger problems, unless covariate gridding on the covariate domain is performed. The Fourier parameterisation, utilising bases with global support (compared to Spline and Gaussian Process basis functions whose support is local on the covariate domain), is generally somewhat more difficult to estimate well in practice, showing greater instability to choices such as starting solution for maximum likelihood estimation. The Constant parameterisation performs surprisingly well in estimating the omnidirectional return values distribution in some cases, but is generally very poor in estimating directional variation. \par

Various choices of methods of inference are also available. Competing approaches include maximum (penalised) likelihood optimisation and Bayesian inference using Markov chain Monte Carlo sampling. It appears however that the major difference, in terms of practical value of inference, is not between frequentist and Bayesian paradigms but rather the advantage gained by exploiting knowledge of likelihood gradient and curvature. In addition, it appears that inference schemes which sample from a negative log likelihood surface randomly, rather that seeking its minimum deterministically, are more stable, and therefore more routinely implementable and useable. Moreover, Bayesian inference gives a more intuitive framework for statistical learning and communication of uncertainty, particularly to a non-specialist audience.

Here, we have focussed on the estimation of non-stationary shape and scale parameters for the conditional distribution of independent peaks over threshold. In practical application, careful estimation of non-stationary extreme value threshold is at least as important for reliable inference. We emphasise that for practical application, any non-stationarity of extreme value threshold should be examined and identified either before or alongside non-stationarity of GP parameters. Anderson et al. (2001) notes that the combination of non-stationary threshold and stationary GP shape and scale is sufficient for modelling a sample of significant wave heights in the North Sea. Physical and statistical intuition suggest, when considering a an extreme value model for a quantity such $H_S$, that non-stationary estimates should be sought in order for each of (a) extreme value threshold, (b) then GP scale, and (c) finally GP shape, and adopted only if justified statistically. Note however that physically plausible oceanographic examples corresponding, for instance, to stationary extreme value threshold but non-stationary GP parameters are also conceivable. In the current work, we assume effectively that any non-stationarity in extreme value threshold as already been identified perfectly and its effect removed from the sample cases considered.

It appears that specification of a piecewise model for the whole sample (e.g. \citealt{BhrEA04}, \citealt{McDEA11}, \citealt{RndEA15a}) incorporating the appropriate tail form, rather than a model for threshold exceedances in isolation is a promising route, since then threshold and tail parameters can be estimated together. However, even the simplest form of a whole sample model requires estimation of additional parameters for the model below the threshold; these parameters are also typically non-stationary with respect to covariates. Efficient and reliable non-stationary estimation using parameterisations and inference schemes similar to those presented here is key.

\section*{Acknowledgement}
We are grateful to Kathryn Turnbull of Lancaster University, UK for useful comments.

\section*{Appendix}

\section*{Maximum likelihood estimation}
%
For maximum likelihood estimation (MLE), we use a back-fitting (or iteratively re-weighted least-squares, IRLS) algorithm to estimate vectors of basis coefficients $\prp{\gpr{}}$ ($\gpr{} = \gpsh{},\gpnu{}$) derived in \citet{JONA2014}. For fixed value of smoothness parameter $\prpn{\gpr{}}$, we initialise coefficients to starting value $\mlsm{\gpr{}}{0}$, and then iterate the following step until convergence

\[ \mlsm{\gpr{}}{i+1} = \left({}\prbsn{\gpr{}}\T{} \enskip \irhs{\mlsm{\gpr{}}{i}} \enskip  \prbsn{\gpr{}} + \prpn{\gpr{}} \prcr{\gpr{}}\right)^{-1}\left({}\prbsn{\gpr{}}\T{} \enskip \irgr{\mlsm{\gpr{}}{i}} + \prbsn{\gpr{}}\T{} \enskip  \irhs{\mlsm{\gpr{}}{i}} \enskip  \prbsn{\gpr{}}\mlsm{\gpr{}}{i}\right){} \]
where
\[ \irgr{\prp{\gpr{}}} = \nabla_{\gpr{}}\mllk{\gpdt{}|\mlpr{}} \text{   and   } \irhs{\prp{\gpr{}}} = -\nabla_{\gpr{}}\nabla_{\gpr{}}\T{}\mllk{\gpdt{}|\mlpr{}} \]
are derived at the end of this Appendix. This algorithm is similar to the mMALA algorithm (see below) used to generate proposals for the corresponding Metropolis-Hastings step in MCMC, in that both exploit first- and second-derivative information to move towards regions of high probability. The back-fitting iteration is of course deterministic, whereas the mMALA step is stochastic. In practice we use the expected values of $\irhs{\prp{\gpr{}}}$ for ease of computation.

\section*{MCMC sampling algorithms}
%
Denoting the set of parameters to be estimated by $\mlpr{} = \{{}\prp{\gpsh{}},\prp{\gpnu{}},\prpn{\gpsh{}},\prpn{\gpnu{}}\}{}$, inference proceeds by sampling from the full conditional distributions $\pd{\mlpr{k}|\gpdt{},\gpdm{},\mlpr{\pexc{}k},\mlhpr{}}$ for each parameter in $\mlpr{}$ in turn, where $\mlhpr{} = \{{}\prpna{\gpsh{}},\prpnb{\gpsh{}},\prpna{\gpnu{}},\prpnb{\gpnu{}}\}{}$ is the set of fixed hyper-parameters for prior distributions. The form of the full conditional distribution varies depending on the type of parameter being estimated, as explained below.

\subsection*{Full conditional distributions for basis coefficients}
%
Vector $\prp{\gpr{}}$ ($\gpr{} = \gpsh{},\gpnu{}$) has the following conditional distribution
\[ \pd{\prp{\gpr{}}|\gpdt{},\gpdm{},\mlpr{\pexc{}\prp{\gpr{}}},\mlhpr{}} \propto{} \pd{\gpdt{}|\gpdm{},\prp{\pexc{}\gpr{}}}\pd{\prp{\gpr{}}|\prpn{\gpr{}}} \]
which is not available in closed form, and therefore cannot be sampled directly in a Gibbs step. Instead, we generate samples using the Metropolis-Hastings algorithm: given current state $\mlsm{\gpr{}}{i}$, we propose new parameter $\mlnw{\gpr{}}$ from proposal distribution
$\pd{\mlnw{\gpr{}}|\mlsm{\gpr{}}{i}}$ and evaluate the acceptance ratio
\[ \mhar{\mlnw{\gpr{}},\mlsm{\gpr{}}{i}} = \frac{\pd{\mlnw{\gpr{}}|\gpdt{},\gpdm{},\mlpr{\pexc{}\prp{\gpr{}}}}\pd{\mlsm{\gpr{}}{i}|\mlnw{\gpr{}}}}{\pd{\mlsm{\gpr{}}{i}|\gpdt{},\gpdm{},\mlpr{\pexc{}\prp{\gpr{}}}}\pd{\mlnw{\gpr{}}|\mlsm{\gpr{}}{i}}} \]
accepting the proposal with probability $q = \min{}(1,A)$, setting $\mlsm{\gpr{}}{i+1} = \mlnw{\gpr{}}$. Otherwise we reject the proposal and set $\mlsm{\gpr{}}{i+1} = \mlsm{\gpr{}}{i}$. As outlined in Section \ref{Sct2.3:Inf} and detailed below, we consider two different methods for generating multivariate proposals for $\prp{\gpr{}}$. In the first approach (referred to as MH), we make an entirely stochastic Gaussian random walk proposal using a fixed covariance matrix; in the second (referred to as mMALA), we make a proposal which is partly deterministic and partly stochastic, accounting for local curvature of the likelihood surface.

For Metropolis-Hastings (MH) inference, we generate Gaussian random walk proposals of the form
\[ \mlnw{\gpr{}} = \mlsm{\gpr{}}{i} + (\prbsn{\gpr{}}\T{}\prbsn{\gpr{}} + \kappa_{\gpr{}} \prcr{\gpr{}})^{-1} \mlep{\gpr{}} \epsilon\]
where $\epsilon$ is a vector of independent standard Normal random variables, and the values of step size $\mlep{\gpr{}}$ and scale factor $\kappa_{\gpr{}}$ are adjusted to achieve reasonable acceptance rates of approximately 0.25. For inference using the Riemann manifold Metropolis-adjusted Langevin algorithm (mMALA, as implemented by \citealt{GIRO2011}) we propose using derivatives of the target distribution at the current sample. This promotes proposals in regions of higher probability, at the additional computational cost of computing necessary derivatives and matrix inverses. At iteration $i$ of the sampling algorithm, where the current sample of the coefficients is $\mlsm{\gpr{}}{i}$, proposals are made as
\[ \mlnw{\gpr{}} = \mlsm{\gpr{}}{i} + \frac{\mlep{\gpr{}}^{2}}{2} \enskip \mlhs{-1}{\mlsm{\gpr{}}{i}} \enskip \mlgr{\mlsm{\gpr{}}{i}} + \mlep{\gpr{}}\sqrt{\mlhs{-1}{\mlsm{\gpr{}}{i}}} \enskip \epsilon \]
where $\epsilon$ is a vector of independent standard Normal random variables, $\mlep{\gpr{}}$ is (adjustable) step size, and
\[ \mlgr{\mlsm{\gpr{}}{i}} = \nabla_{\prp{\gpr{}}}\mllk{\prp{\gpr{}}} \bigg|_{\mlsm{\gpr{}}{i}} \text{   and   }  \mlhs{}{\mlsm{\gpr{}}{i}} = -\nabla_{\prp{\gpr{}}}\nabla_{\prp{\gpr{}}}\T{}\mllk{\prp{\gpr{}}} \bigg|_{\mlsm{\gpr{}}{i}} \]
are the negative gradient and negative Hessian of the log density, with
\[\mllk{\prp{\gpr{}}} = \log{} \pd{\prp{\gpr{}}|\gpdt{},\gpdm{},\mlpr{\pexc{}\prp{\gpr{}}},\mlhpr{}} \text{   and   } \nabla_{\prp{\gpr{}}} = (\partial{}/\partial{}\prp{\eta{}1},\ldots{},\partial{}/\partial{}\prp{\eta{}\prnbs{}})\T{}  \text{   .   }\]
Computation of likelihood derivatives is described at the end of this Appendix. In practice we use the expected values of $\mlhs{}{\mlsm{\gpr{}}{i}}$ for ease of computation.

\subsection*{Full conditional distributions for prior precisions}

Prior precision parameter $\prpn{\gpr{}}$ ($\gpr{} = \gpsh{},\gpnu{}$) has the following conditional distribution
\[ \pd{\prpn{\gpr{}}|\gpdt{},\gpdm{},\mlpr{\pexc{}\prpn{\gpr{}}},\mlhpr{}} \propto{} \pd{\prp{\gpr{}}|\prpn{\gpr{}}}\pd{\prpn{\gpr{}}|\prpna{\gpr{}},\prpnb{\gpr{}}} \text{   .   }\]
By construction, since the Gamma distribution is a conjugate prior for the precision of a Gaussian distribution, we know that the full conditional distribution is also Gamma, with updated parameters
\[ \gba{\gpr{}} = \prpna{\gpr{}} + \frac{\prnbs{\gpr}}{2} \text{   and   } \gbb{\gpr{}} = \prpnb{\gpr{}} + \frac{1}{2} \prp{\gpr{}}\T{}\prcr{\gpr{}}\prp{\gpr{}} \text{   .   }\]

\section*{Derivatives of the posterior distribution}

Here we find the derivatives of the log posterior distribution, required for maximum likelihood and mMALA inference. The log likelihood of the observed data under the generalised Pareto distribution is
\begin{equation*}
\mllk{\gpdt{}|\mlpr{}} = \begin{cases}
\sum_{i=1}^{\gpnd{}} {\big[{}-\log{}\left({}\frac{\gpnu{i}}{1+\gpsh{i}}\right){} - \left({}\frac{1}{\gpsh{i}}+1\right){}\log{}\left({}1+\frac{\gpsh{i}}{\gpnu{i}}(1+\gpsh{i})\gpdt{i}\right){}\big]{}} & \text{ for } \gpsh{i} \neq{} 0 \\
\sum_{i=1}^{\gpnd{}} \big[{}-\log{}\left({}\frac{\gpnu{i}}{1+\gpsh{i}}\right){} - \frac{(1+\gpsh{i})\gpdt{i}}{\gpnu{i}}\big]{} & \text{ for } \gpsh{i} = 0 \text{   .   }
\end{cases}
\end{equation*}
The log conditional distribution for the vector of basis coefficients $\prp{\gpr{}}$ ($\gpr{} = \gpsh{},\gpnu{}$) is then the sum of this likelihood plus a contribution from the prior distribution
\begin{align*}
\mllk{\prp{\gpr{}}} & = \log{}\left({}\pd{\prp{\gpr{}}|\gpdt{},\gpdm{},\mlpr{\pexc{}\prp{\gpr{}}},\mlhpr{}}\right){} \\
& = \mllk{\gpdt{}|\mlpr{}} - \frac{\prpn{\gpr{}}}{2}\prp{\gpr{}}\T{}\prcr{\gpr{}}\prp{\gpr{}} \text{   .   }
\end{align*}
We note the equivalence between this expression and the penalised (negative log) likelihood used for maximum likelihood inference. The gradient of the log conditional distribution for vector of coefficients $\prp{\gpr{}}$ ($\gpr{} = \gpsh{},\gpnu{}$) is
\begin{equation*}
\nabla_{\prp{\gpr{}}}\mllk{\prp{\gpr{}}} = \nabla_{\prp{\gpr{}}}\mllk{\gpdt{}|\mlpr{}} - \prpn{\gpr{}} \prcr{\gpr{}}\prp{\gpr{}} \text{   .   }
\end{equation*}
Using the chain rule, the likelihood gradient can be computed as
\begin{align*}
\nabla_{\prp{\gpr{}}}\mllk{\gpdt{}|\mlpr{}} & = \left( {}\nabla_{\prp{\gpr{}}} (\prbsn{\gpr{}}\prp{\gpr{}})\right){}\T{}\left( {}\nabla_{\gpr{}} \mllk{\gpdt{}|\mlpr{}}\right){} \\
& = \prbsn{\gpr{}}\T{}\left({}\nabla_{\gpr{}}\mllk{\gpdt{}|\mlpr{}}\right){} \text{   .   }
\end{align*}

The components of $\nabla_{\gpsh{}}\mllk{\gpdt{}|\mlpr{}}$ are computed as
\begin{equation}
\frac{\partial{}}{\partial{}\gpsh{i}}\mllk{\gpdt{}} = \begin{cases}
-\frac{1}{\gpsh{i}^{2}G_{i}}(1-2\gpsh{i})(G_{i}-1) + \frac{1}{1+\gpsh{i}} + \frac{1}{\gpsh{i}}\log{}(G_{i}) & \text{ for } \gpsh{i} \neq{} 0 \\
-\frac{\gpdt{i}}{\gpnu{i}} + \frac{1}{1+\gpsh{i}} & \text{ for } \gpsh{i} = 0
\end{cases} \nonumber
\end{equation}
where $G_{i} = 1 + \frac{\gpsh{i}}{\gpnu{i}}(1+\gpsh{i})\gpdt{i}$, and the components of $\nabla_{\gpnu{}}\mllk{\gpdt{}|\mlpr{}}$ are
\begin{equation*}
\frac{\partial{}}{\partial{}\gpnu{i}}\mllk{\gpdt{}} = \begin{cases}
\frac{1}{\gpnu{i}}\left({}1 - \big({}\frac{1}{\gpsh{i}}+1\big){}\frac{G_{i}-1}{G_{i}}\right){} & \text{ for } \gpsh{i} \neq{} 0 \\
\frac{1}{\gpnu{i}}\left({}1 - \frac{G_{i}-1}{\gpsh{i}}\right){} & \text{ for } \gpsh{i} = 0 \text{   .   }
\end{cases}
\end{equation*}
Differentiating $\nabla_{\prp{\gpr{}}}\mllk{\prp{\gpr{}}}$ ($\gpr{} = \gpsh{},\gpnu{}$) again gives the Hessian matrix
\begin{equation*}
\nabla_{\prp{\gpr{}}}\nabla_{\prp{\gpr{}}}\T{}\mllk{\prp{\gpr{}}} = \nabla_{\prp{\gpr{}}}\nabla_{\prp{\gpr{}}}\T{}\mllk{\gpdt{}|\mlpr{}} - \prpn{\gpr{}} \prcr{\gpr{}} \text{   .   }
\end{equation*}
Applying the chain rule
\begin{equation*}
\nabla_{\prp{\gpr{}}}\nabla_{\prp{\gpr{}}}\T{}\mllk{\gpdt{}|\mlpr{}} = \prbsn{\gpr{}}\T{}\left({}\nabla_{\gpr{}}\nabla_{\gpr{}}\T{}\mllk{\gpdt{}|\mlpr{}}\right){}\prbsn{\gpr{}} \text{   .   }
\end{equation*}
Note that the components of $\nabla_{\gpr{}}\mllk{\gpdt{}|\mlpr{}}$ and $(\nabla_{\gpr{}}\nabla_{\gpr{}}\T{}\mllk{\gpdt{}|\mlpr{}})$ are computed separately for $\gpr{} = \gpsh{}$ and $\gpr{} = \gpnu{}$. Further, the expected values of likelihood second derivatives with respect to $\gpsh{}$ and $\gpnu{}$ are
\begin{equation*}
-\mathbb{E}_Y \left[\frac{\partial^{2}}{\partial{}\gpsh{i}\partial{}\gpsh{j}}\mllk{\gpdt{}|\mlpr{}}\right] = \begin{cases}
\frac{1}{(1+\gpsh{i})^{2}} & \text{ for } i = j \\
0 & \text{ for } i \neq{} j
\end{cases}
\end{equation*}
and
\begin{equation*}
-\mathbb{E}_Y \left[\frac{\partial^{2}}{\partial{}\gpnu{i}\partial{}\gpnu{j}}\mllk{\gpdt{}|\mlpr{}}\right] = \begin{cases}
\frac{1}{\gpnu{}^{2}(1+2\gpsh{i})} & \text{ for } i = j \\
0 & \text{ for } i \neq{} j
\end{cases}
\end{equation*}
such that Hessian matrices are diagonal. Moreover, the expectations of all of the cross derivatives $\frac{\partial^{2}}{\partial{}\gpsh{i}\partial{}\gpnu{j}}\mllk{\gpdt{}|\mlpr{}}$ are zero, since estimates of $\gpsh{}$ and $\gpnu{}$ are asymptotically independent by construction (e.g. \citealt{ChvDvs05}).

\bibliographystyle{elsarticle-harv}
\section*{References}
\bibliography{SmoothingBib}

\end{document}